%% file: SMEFT-RW.tex
\documentclass[submission, Phys]{SciPost}
\pdfoutput=1

\numberwithin{equation}{section}
\numberwithin{figure}{section}
\numberwithin{table}{section}

\newcommand{\be}{\begin{equation}}
\newcommand{\ee}{\end{equation}}
\newcommand{\la}{\left\langle}
\newcommand{\ra}{\right\rangle}
\newcommand{\lc}{\left[}
\newcommand{\rc}{\right]}
\newcommand{\lp}{\left(}
\newcommand{\rp}{\right)}

\usepackage{amssymb}
\usepackage{amsmath}
\usepackage{booktabs}

\begin{document}

\begin{flushright}
Nikhef-2019-019,
OUTP-19-06P
\end{flushright}

\begin{center}{\Large \textbf{
Constraining the SMEFT with Bayesian reweighting
}}\end{center}

\begin{center}
Samuel van Beek,\textsuperscript{1}
Emanuele R. Nocera,\textsuperscript{1*}
Juan Rojo,\textsuperscript{1,2}
and Emma Slade\textsuperscript{3}
\end{center}

\begin{center}
{\bf 1} Nikhef Theory Group, Science Park 105, 1098 XG Amsterdam, The
  Netherlands.\\[0.1cm]
  {\bf 2}  Department of Physics and Astronomy, Vrije Universiteit, 1081 HV
  Amsterdam. \\[0.1cm]
 {\bf 3} Rudolf Peierls Centre for Theoretical Physics, University of Oxford,\\ 
 Clarendon Laboratory, Parks Road, Oxford OX1 3PU, United Kingdom
\\[0.1cm]
*\href{mailto:e.nocera@nikhef.nl}{e.nocera@nikhef.nl}
\end{center}


\section*{Abstract}
{\bf We illustrate how Bayesian reweighting can be used to incorporate the 
  constraints provided by new measurements into a global Monte Carlo analysis
  of the Standard Model Effective Field Theory (SMEFT).
  This method, extensively applied to study the impact of new data
  on the parton distribution functions of the proton, is here validated 
  by means of our recent SMEFiT analysis of the top quark sector.
  We show how, under well-defined conditions and for the SMEFT operators 
  directly sensitive to the new data, the reweighting procedure is equivalent 
  to a corresponding new fit.
  We quantify the amount of information added to the SMEFT parameter space 
  by means of the Shannon entropy and of the Kolmogorov-Smirnov statistic.
  We investigate the dependence of our results upon the choice of 
  alternative expressions of the weights.
}

\vspace{5pt}
\noindent\rule{\textwidth}{1pt}
\tableofcontents\thispagestyle{fancy}
\noindent\rule{\textwidth}{1pt}
\vspace{5pt}

\input{sec-introduction.tex}

\input{sec-rw.tex}

\input{sec-validation.tex}

\input{sec-gk.tex}

\input{sec-summary.tex}

\section*{Acknowledgments}
We are grateful to Stefano Forte for discussions and suggestions
related to the results of this paper.
S.~v.~B. is grateful for the hospitality of the Rudolf Peierls
Centre for Theoretical Physics at the University of Oxford
where part of this work was carried out.

\paragraph{Funding information}
J.~R. and E.~S. are supported by the European Research Council Starting
Grant ``PDF4BSM''.
J.~R. is also partially supported by the Netherlands Organisation for 
Scientific Research (NWO).
E.~R.~N. is supported by the European Commission through the Marie 
Sk\l{}odowska-Curie Action ParDHonS FFs.TMDs (grant number 752748).

\appendix

\input{sec-appendix}

\bibliography{SMEFT-RW}

\nolinenumbers

\end{document}

%% file: sec-introduction.tex
\section{Introduction}
\label{sec:sec-intro}

A powerful framework to parametrise and constrain potential
deviations from Standard Model (SM) predictions in a model-independent 
way is provided by the SM Effective Field Theory
(SMEFT)~\cite{Weinberg:1979sa,Buchmuller:1985jz,Grzadkowski:2010es}, 
see~\cite{Brivio:2017vri} for a recent review.
In the SMEFT, effects of beyond the SM (BSM) dynamics at high scales 
$E\simeq \Lambda$ are parametrised, for $E\ll \Lambda$, in terms of 
higher-dimensional operators built up from the SM fields and symmetries.
This approach is fully general, as one can construct complete bases of 
independent operators, at any given mass dimension, that 
can be systematically matched to ultraviolet-complete theories.

Analysing experimental data in the SMEFT framework, however, is far
from straightforward because of the large dimensionality of the underlying 
parameter space.
For instance, without flavour assumptions, one needs to deal with 
$N_{\rm op}=2499$ independent operators corresponding to three 
fermion generations.
Because of this challenge, the complexity and breadth of SMEFT analyses, in 
particular of LHC data, has been restricted to a subset of higher-dimensional 
operators so far, typically clustered in sectors that are assumed to be 
independent from each other~\cite{Alioli:2017jdo,Englert:2015hrx,Ellis:2014jta,
Alioli:2017nzr,Alioli:2018ljm,Alte:2017pme,AguilarSaavedra:2018nen,
Castro:2016jjv,deBlas:2016ojx,Ellis:2018gqa,Ellis:2014dva,Buckley:2016cfg,
Butter:2016cvz,Azatov:2015oxa,Biekotter:2018rhp,
Schulze:2016qas,Cirigliano:2016nyn,Alioli:2017ces}.

More recently, some of us have developed a novel approach to efficiently 
explore the parameter space in a global analysis of the SMEFT: 
the SMEFiT framework~\cite{Hartland:2019bjb}.
This approach is inspired by the NNPDF
methodology~\cite{Ball:2008by,Ball:2010de,Ball:2012cx,Ball:2014uwa,Ball:2017nwa}
for the determination of the parton distributions functions (PDFs) of the 
proton~\cite{Gao:2017yyd,Forte:2013wc,Kovarik:2019xvh}.
The SMEFiT methodology realises a Monte Carlo representation of the probability 
distribution in the space of the SMEFT parameters, whereby each parameter is 
associated to a statistical ensemble of equally probable replicas.
Two of the main strengths of this framework are the ability to deal
with arbitrarily large or complicated parameter spaces, and to avoid any 
restriction on the theory calculations used, {\it e.g.} in relationship with
the inclusion of higher-order EFT terms.
As a proof of concept, SMEFiT was used
in~\cite{Hartland:2019bjb} to analyse about a hundred
top quark production measurements from the LHC.
In total, $N_{\rm op}=34$ independent degrees of freedom at mass-dimension six 
were constrained simultaneously,
including both linear, $\mathcal{O}\lp \Lambda^{-2}\rp$, and quadratic, 
$\mathcal{O}\lp \Lambda^{-4}\rp$, EFT effects as well as NLO QCD corrections.

As more experimental data becomes available, the probability distribution in
the space of the SMEFT parameters should be correspondingly updated.
This can naturally be achieved by performing a new fit to the extended set of 
data, which will however require in general a significant
computational effort.
In many situations, however, one would like to quantify the impact of a 
new measurement to the SMEFT parameter space more efficiently, {\it i.e.} 
without having to perform an actual fit.
This may routinely happen whenever a new measurement is presented by the
LHC experimental collaborations.
In order to do so, one may wonder whether methods developed in order to 
quantify the PDF sensitivity to new data can help, such as the profiling of 
Hessian PDF sets~\cite{Paukkunen:2014zia,Eskola:2019dui} or the Bayesian
reweighting of Monte Carlo PDF sets~\cite{Ball:2010gb,Ball:2011gg}.

The aim of this work is to demonstrate that Bayesian reweighting, originally
developed for Monte Carlo PDF sets in~\cite{Ball:2010gb,Ball:2011gg}, can 
be successfully extended to the SMEFiT framework.
We do this as a proof of concept: given a prior SMEFiT fit based on a variant
of our previous study~\cite{Hartland:2019bjb}, we show that single top-quark 
production measurements can be equivalently included in the prior
either by Bayesian reweighting or by a new fit. 
We quantify the amount of new information that the measurements are 
bringing into the SMEFT parameter space by means of appropriate
estimators, such as the Shannon entropy and the Kolmogorov-Smirnov statistic.
We discuss the limitations of the method, explore the conditions under 
which it can be safely applied, and study its dependence upon a different
definition of the replica weights, as proposed by Giele and 
Keller~\cite{Giele:1998gw,Giele:2001mr} (see also~\cite{Sato:2013ika}).

The outline of this paper is as follows.
In Sect.~\ref{sec:sec-rw} we review the Bayesian reweighting method in the
context of a SMEFiT analysis.
In Sect.~\ref{sec:sec-validation} we validate the method by reweighting a 
SMEFiT prior with different single-top datasets and by comparing the results
with the corresponding fits.
In Sect.~\ref{sec:sec-gk} we study the sensitivity of the reweighting method
upon alternative definitions of the weights.
We summarise our findings in Sect.~\ref{sec:sec-summary}.
Our results are made publicly available in the form of a stand-alone
{\tt Python} code, which we describe in the Appendix.

%% file: sec-rw.tex
\section{Bayesian reweighting revisited}
\label{sec:sec-rw}

Bayesian reweighting was originally developed in the case of 
PDFs in Refs~\cite{Ball:2010gb,Ball:2011gg}, inspired
by the earlier studies of~\cite{Giele:1998gw,Giele:2001mr}.
It assumes that the probability density in the space of PDFs is represented by
an ensemble of $N_{\rm rep}$ equally probable Monte Carlo replicas
\be
\label{eq:probdensity}
\{\,f_i^{(k)}(x,Q_0)\,\} \, ,\quad i=1,\ldots,n_f \, ,
\quad k=1,\ldots,N_{\rm rep} \, ,
\ee
where $n_f$ is the number of active partons at the initial parametrisation 
scale $Q_0$ and $f$ is the corresponding PDF.
The ensemble, Eq.~(\ref{eq:probdensity}), is obtained by sampling $N_{\rm rep}$ 
replicas from the experimental data and then by performing a separate PDF fit 
to each of them.

Analogously, a SMEFiT analysis represents the probability density in 
the space of Wilson coefficients (or SMEFT parameters) 
as an ensemble of $N_{\rm rep}$  Monte Carlo replicas
\be
\label{eq:samplingSMEFT}
\{\, c_i^{(k)} \, \} ,\quad i=1,\ldots,N_{\rm op} \, ,
\quad k=1,\ldots,N_{\rm rep} \, ,
\ee
where $N_{\rm op}$ is the number of independent dimension-6 operators
$\{\mathcal{O}_i^{(6)}\}$ that define the fitting basis of the analysis 
and $c_i$ are the corresponding Wilson coefficients that enter the SMEFT
Lagrangian,
\begin{equation}
\label{eq:smeftlagrangian}
\mathcal{L}_{\rm SMEFT}=\mathcal{L}_{\rm SM} 
+ 
\sum_i^{N_{\rm op}} \frac{c_i}{\Lambda^2}\mathcal{O}_i^{(6)} \, ,
\end{equation}
with $\Lambda$ the characteristic energy scale where new physics sets in.
Since we neglect operator running effects~\cite{Jenkins:2013zja}, 
the coefficients $\{c_i^{(k)}\}$ are scale independent.
The ensemble, Eq~\eqref{eq:samplingSMEFT}, can be obtained as a result of 
a fit in the SMEFiT framework.

The starting point of Bayesian reweighting is therefore a realisation 
of Eq.~(\ref{eq:samplingSMEFT}), which we will henceforth call the {\it prior}.
The next step is to quantify the impact of some new measurement on the prior.
Following Bayesian inference, this can be achieved by associating a weight 
$\omega_k$ to each Monte Carlo replica in the prior.
The value of these weights depends on the agreement (or lack thereof) between
the theory predictions constructed from each replica in the prior and the new 
dataset.
Their analytic expression is~\cite{Ball:2010gb,Ball:2011gg}
\be
\label{eq:weightsRW}
\omega_k 
\propto 
\left(\chi^2_k \right)^{(n_{\rm dat}-1)/2}
\,
\exp\left( -\chi^2_k/2\right) \, ,\quad k=1,\ldots, N_{\rm rep} \, ,
\ee
where $n_{\rm dat}$ is the number of data points in the new dataset and
$\chi^2_k$ is the unnormalised $\chi^2$ of the new dataset computed 
with the $k$-th replica in the prior.
These weights are normalised in such a way that their sum adds up to
the total number of replicas, namely
\be
\label{eq:normalisation}
\sum_{k=1}^{N_{\rm rep}} \omega_k= N_{\rm rep}\, .
\ee
We will discuss in Sect.~\ref{sec:sec-gk} how results are affected if the
Giele-Keller expression of the weights~\cite{Giele:1998gw,Giele:2001mr}, 
which differs from Eq.~\eqref{eq:weightsRW}, is used instead.
After the inclusion of the new data, replicas are no longer equally probable.
The statistical features of the ensemble should therefore be computed 
accordingly.
For instance, the new expectation values are given by weighted means
\be
\label{eq:weighted_ave}
\la c_i\ra = \frac{1}{N_{\rm rep}}\sum_{k=1}^{N_{\rm rep}}\,\omega_k\,c_i^{(k)} \, ,
\ee
and likewise for other estimators such as variances and correlations.


In the definition of the weights, Eq.~(\ref{eq:weightsRW}),
the unnormalised $\chi^2_k$ associated to the $k$-th
Monte Carlo replica is constructed as
\begin{equation}
  \chi^2_k = \sum_{i,j=1}^{n_{\rm dat}}\lp 
  \mathcal{F}^{(\rm th)}_i\lp \{  c_k \} \rp
  -\mathcal{F}^{(\rm exp)}_i\rp ({\rm cov}^{-1})_{ij}
\lp 
  \mathcal{F}^{(\rm th)}_j\lp \{ c_k \} \rp
  -\mathcal{F}^{(\rm exp)}_j\rp
 \label{eq:chi2definition2}
    \; ,
\end{equation}
where $\mathcal{F}^{(\rm th)}_i\lp \{  c_k \} \rp$ is the theoretical prediction
for the $i$-th cross section $\mathcal{F}_i$ evaluated using the Wilson 
coefficients associated to the k-th replica, $\{  c_k \} $, 
and $\mathcal{F}^{(\rm exp)}_i$
is the central value of the corresponding experimental measurement.
Note that in Eq.~(\ref{eq:chi2definition2}) the sum runs over only the data 
points of the new dataset that is being added by reweighting, while all the
information from the prior fit is encoded in the Wilson coefficients 
$\{ c_k \} $ associated to the corresponding Monte Carlo sample.

The total covariance 
matrix, ${\rm cov}_{ij}$ in Eq.~(\ref{eq:chi2definition2}), should contain all the relevant sources of
experimental and theoretical uncertainties.
Assuming that theoretical uncertainties follow an underlying Gaussian
distribution, and that they are uncorrelated
to the experimental uncertainties, it can be shown~\cite{Ball:2018odr} that 
\be
\label{eq:covmatsplitting}
{\rm cov}_{ij} = {\rm cov}^{(\rm exp)}_{ij} + {\rm cov}^{(\rm th)}_{ij} \, ,
\ee
that is, the total covariance matrix
is given by the sum of the experimental and theoretical covariance matrices.
The experimental covariance matrix is constructed
using the `$t_0$' prescription~\cite{Ball:2009qv},
\begin{eqnarray}
  ({\rm cov_{t_0}})^{(\rm exp)}_{ij}\equiv
  \lp \sigma^{\rm (stat)}_i\rp^2\delta_{ij}
  &+&
  \Bigg(\sum_{\alpha=1}^{N_{\rm sys}}\sigma_{i,\alpha}^{\rm (sys)}
  \sigma_{j,\alpha}^{\rm (sys)}\mathcal{F}_{i}^{(\rm exp)}\mathcal{F}_{j}^{(\rm exp)} \nonumber\\
  &+&\sum_{\beta=1}^{N_{\rm norm}}\sigma_{i,\beta}^{\rm (norm)}
  \sigma_{j,\beta}^{\rm (norm)}\mathcal{F}_{i}^{(\rm th,0)}\mathcal{F}_{j}^{(\rm th,0)}\Bigg)\; ,
  \label{eq:t0covmat}
\end{eqnarray}
where `sys' (`norm') indicates the additive (multiplicative) relative experimental systematic errors separately; $\mathcal{F}_{i}^{(\rm th,0)}$
corresponds to
a fixed set of theoretical  predictions obtained from a previous fit.
All available sources of statistical and systematic uncertainties
for a given dataset are considered in Eq.~(\ref{eq:t0covmat}), including 
bin-by-bin correlations whenever available.
The theoretical covariance matrix includes only the contribution from the
PDF uncertainties in this analysis. It is given by
\be
\label{eq:PDFcovmat}
{\rm cov}^{(\rm th)}_{ij} = \la 
\mathcal{F}^{(\rm th)(r)}_i \mathcal{F}^{(\rm th)(r)}_j
\ra_{\rm rep} -
\la 
\mathcal{F}^{(\rm th)(r)}_i \ra_{\rm rep} \la\mathcal{F}^{(\rm th)(r)}_j
\ra_{\rm rep} \, ,
\ee
where the theoretical predictions $\mathcal{O}^{(\rm th)(r)}_i$
are computed using the SM theory and the $r$-th replica
from the NNPDF3.1NNLO no-top PDF set (which excludes all top quark measurements
to avoid double counting).


After reweighting, replicas with small weights become almost irrelevant.
This implies that the reweighted ensemble will be less efficient 
than the prior in representing the probability distribution in the 
space of SMEFT parameters.
This loss of efficiency is quantified by the Shannon entropy (or the effective 
number of replicas left after reweighting)
\be
\label{eq:neff}
N_{\rm eff}
=
\exp \left( \frac{1}{ N_{\rm rep}} \sum_{k=1}^{N_{\rm rep}} \omega_k \ln \frac{N_{\rm rep}}{\omega_k}\right) \, .
\ee
This is the number of replicas needed in a hypothetical new fit to obtain an 
ensemble as accurate as the reweighted ensemble.
If $N_{\rm eff}$ becomes too low, the reweighting procedure no longer 
provides a reliable representation of the probability distribution in the 
space of SMEFT parameters.
As a rule of thumb, in this work we require $N_{\rm eff}\gtrsim 100$.
This value was determined by studying the dependence of the SMEFiT results
upon the number of Monte Carlo replicas in our previous 
study~\cite{Hartland:2019bjb}. 
If $N_{\rm eff}\lesssim 100$, either a prior set
consisting of a larger number of starting replicas
$N_{\rm rep}$ or a new fit would be required to properly
incorporate the information contained in the new data.

Such a situation can happen in two cases.
First, if the new data contains a lot of new information, for example 
because it heavily constrains a new region of the parameter space or 
because it has a large statistical power.
Second, if the new data is inconsistent with the old in the theoretical
framework provided by the SMEFT.
These two cases can be distinguished by examining the $\chi^2$ profile
of the new data: if there are very few replicas with a $\chi^2$ per data point 
of order unity (or lower) in the reweighted ensemble, then the new data is 
inconsistent with the old.
The inconsistency can be quantified by computing the $\mathcal{P}(\alpha)$ 
distribution (see~\cite{Ball:2010gb} for further details), defined as 
\be
\label{eq:Pofalpha}
\mathcal{P}(\alpha)
\propto
\frac{1}{\alpha}\sum_{k=1}^{N_{\rm rep}}\omega_k(\alpha)\, ,
\ee
where $\alpha$ is the factor by which the uncertainty on the new data must be 
rescaled to make them consistent with the old.
If $\mathrm{argmax}\,\mathcal{P}(\alpha)\sim 1$, the new data is consistent 
with the old; if $\mathrm{argmax}\,\mathcal{P}(\alpha)\gg1$, it is not.

A limitation of the effective number of replicas $N_{\rm eff}$, 
Eq.~\eqref{eq:neff}, and of the $\mathcal{P}(\alpha)$ distribution,
Eq.~\eqref{eq:Pofalpha}, is that they provide only a global measure 
of the impact of the new data.
They do not allow one to determine which specific directions
of the SMEFT operator space are being constrained the most.
Such an information can be instead accessed by means of the  
Kolmogorov-Smirnov (KS) statistic.
This estimator is defined as
\be
\label{eq:KS-test}
{\rm KS}
= 
\underset{\langle c_i\rangle}{\text{sup}} 
\large| F_{\text{rw}} (\langle c_i \rangle ) - F_{\text{fit}} (\langle c_i \rangle) \large| \,,
\ee
{\it i.e.} as the supremum of the set of distances between the 
reweighted and the refitted probability distributions for each SMEFT operator,
$F_{\text{rw}} (\langle c_i \rangle)$ and $F_{\text{fit}} (\langle c_i \rangle)$,
respectively.
Clearly $0\leq {\rm KS}\leq 1$: ${\rm KS} \sim 0$
if the coefficients obtained either from 
reweighting or from a new fit belong to ensembles that represent the same 
probability distribution;
${\rm KS}\to 1$ if they belong to ensembles that represent
different probability distributions.

\begin{figure}[!t]
  \begin{center}
    \includegraphics[scale=0.95]{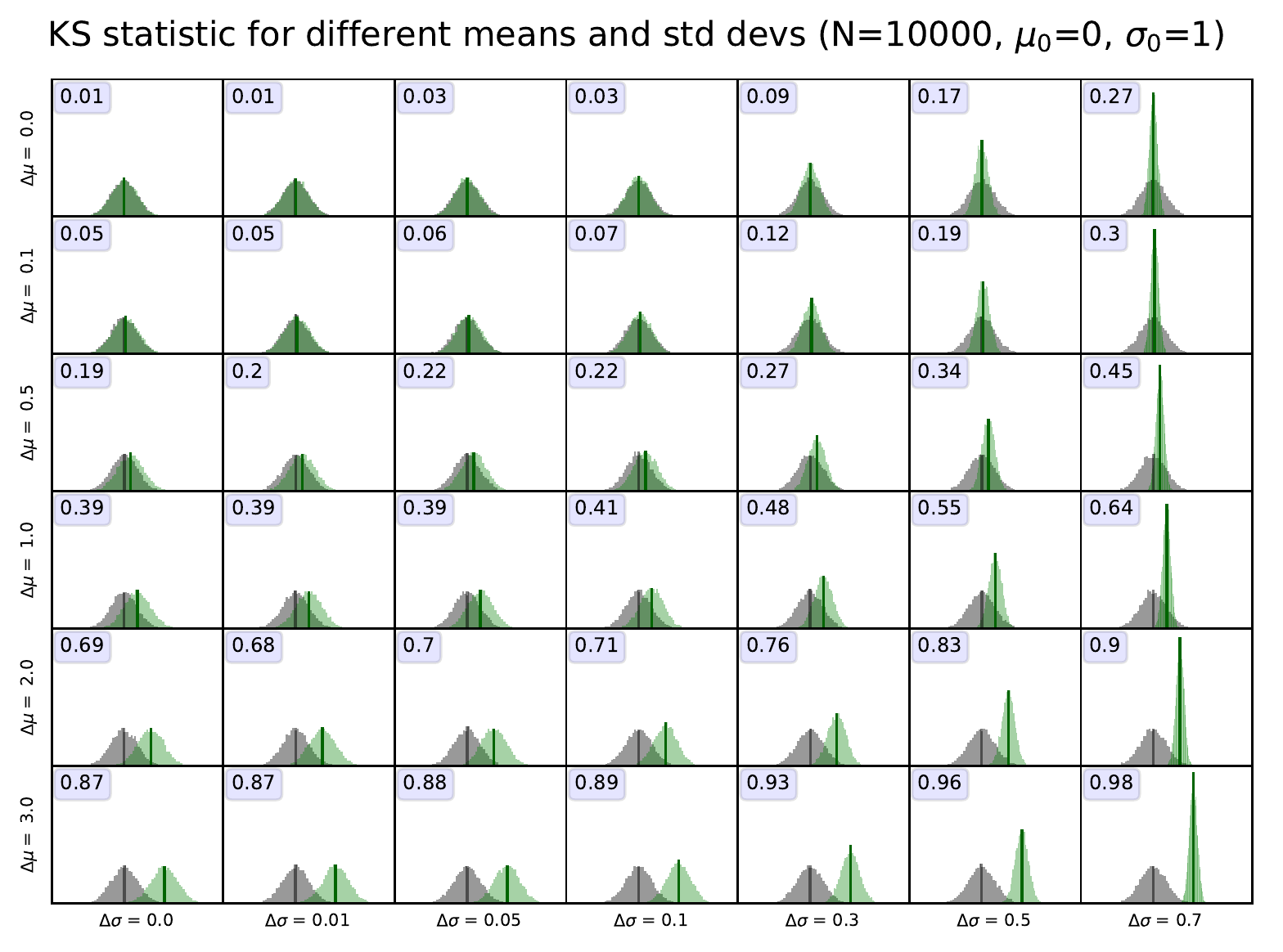}
    \caption{\small The value of the KS statistic, Eq.~\eqref{eq:KS-test},
      between two Gaussian distributions sampled $N_{\rm rep}=10^4$ times each.
      The grey distribution has mean $\mu_0=0$ and standard deviation 
      $\sigma_0=1$. The green distribution has mean $\mu_1=\Delta\mu$ and 
      standard deviation $\sigma_1=\sigma_0-\Delta\sigma$.
     \label{fig:KS-summary} }
  \end{center}
\end{figure}

The transition between the two regimes is smooth.
As an example, in Fig.~\ref{fig:KS-summary} we show the value of the KS 
statistic, Eq.~\eqref{eq:KS-test}, between two Gaussian distributions sampled 
$N_{\rm rep}=10^4$ times each.
One distribution (grey histogram) has mean $\mu_0=0$ and standard deviation 
$\sigma_0=1$, while the other distribution (green histogram) has mean 
$\mu_1=\Delta\mu$ and standard deviation $\sigma_1=\sigma_0-\Delta\sigma$.
The values of $\Delta \mu$ and of $\Delta\sigma$ are being increased from 
top to bottom and from left to right, respectively.
While the KS statistic does not provide a clear-cut threshold to classify the 
two distributions as the same or not, it can be used as a guide to 
disentangle genuine effects of new data in the SMEFT operator space from 
statistical fluctuations.
We should note that the examples shown in Fig.~\ref{fig:KS-summary} are only 
valid for Gaussian distributions; in general, the probability distributions of 
Wilson coefficients can be non-Gaussian.

After reweighting, the prior is accompanied by a set of weights.
For practical reasons, it is convenient to replace both of them with a new
set of replicas which reproduce the reweighted probability distribution in the 
space of SMEFT parameters, but are again equally probable.
This can be achieved by means of unweighting~\cite{Ball:2011gg}.
For statistical purposes, {\it e.g.} for the calculation of 95\% confidence 
level (CL) intervals, the unweighted set can be treated in 
the same way as the prior (and as all sets obtained in a new fit).

%% file: sec-validation.tex
\section{Reweighting the SMEFT parameter space}
\label{sec:sec-validation}

We now explicitly illustrate how Bayesian reweighting works with a SMEFT
Monte Carlo fit.
We first describe our choice of prior for Eq.~\eqref{eq:samplingSMEFT} 
and we reweight and unweight it with several sets of single-top 
production data.
We then monitor the efficiency loss of the reweighted set and we verify under 
which conditions reweighting lead to results equivalent to those of a new fit.
We finally test such conditions upon variation of the process type used to 
reweight the prior.

\subsection{Choice of prior and of reweighting datasets}

We choose the prior for Eq.~(\ref{eq:samplingSMEFT}) as a fit obtained in the 
SMEFiT framework from our previous work~\cite{Hartland:2019bjb}.
Specifically we consider a variant of our baseline result,
where measurements of inclusive single-top quark production in the 
$t$-channel~\cite{CMS-PAS-TOP-14-004,Aaboud:2017pdi,CMS:2016xnv,
Aaboud:2016ymp} and in the
$s$-channel~\cite{Khachatryan:2016ewo,Aad:2015upn} are removed from the 
default dataset: in total $n_{\rm dat}=20$ data points for single-top 
$t$-channel production (total and differential cross sections) 
and $n_{\rm dat}=2$ data points (total cross sections) for single-top 
$s$-channel production.
The prior is thus based on $n_{\rm dat}=81$ data points (the 103 used in 
the baseline fit of~\cite{Hartland:2019bjb} minus the above 22).

To ensure a sufficiently accurate representation of the probability 
distribution in the SMEFT parameter space, the prior is made of 
$N_{\rm rep}=10^4$ Monte Carlo replicas.
Such a large sample -- one order of magnitude larger than the sample used
in~\cite{Hartland:2019bjb} -- is required to mitigate the efficiency loss 
upon reweighting.
The effective number of replicas, Eq.~(\ref{eq:neff}), would otherwise become 
too small and reweighted results will no longer be reliable.

The prior is then reweighted with the datasets of single-top production
listed in Table~\ref{eq:input_datasets2}. 
These sets include all the sets originally removed from the default fit 
in~\cite{Hartland:2019bjb} to generate the prior.
Each of them is labelled as in our previous work 
(see Table 3.3. in~\cite{Hartland:2019bjb}).
In addition, we consider three extra datasets for the total cross sections 
from CMS at 8 and 13 TeV and from ATLAS at 8 TeV (for a total of $n_{\rm dat}=5$ 
data points).
This data was not taken into account in the fit of Ref.~\cite{Hartland:2019bjb} 
to avoid a double-counting issue, since we already included the corresponding 
absolute differential distributions determined from the same data taking.
We believe that this is not an issue here, since our aim is to validate the 
reweighting procedure rather than to extract accurate bounds on the SMEFT 
parameters.
We therefore retain all the datasets collected in 
Table~\ref{eq:input_datasets2}.

\input{tables/input-datasets.tex}

\begin{figure}[!t]
\centering
\includegraphics[width=0.99\textwidth]{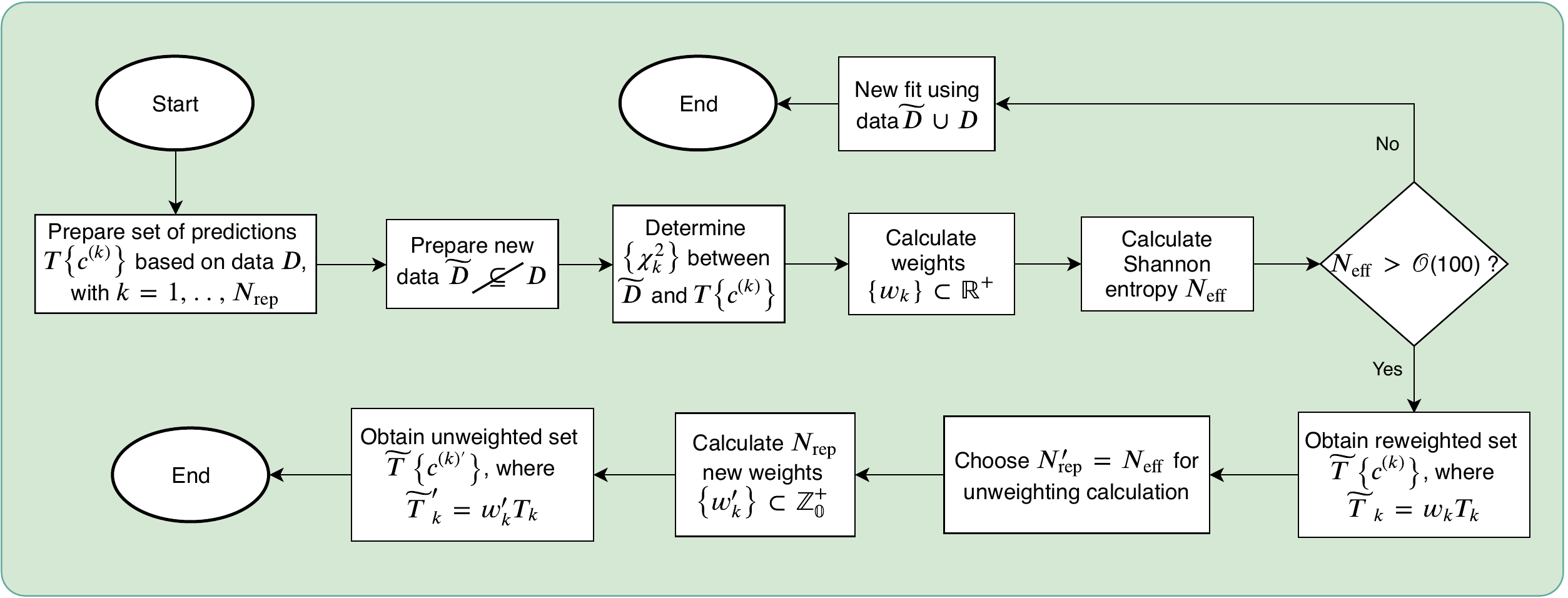}\\
\caption{\small Overview of the reweighting/unweighting procedure.
      The procedure is successful if the probability distribution associated
      to the unweighted set coincides with the one from a new fit.
\label{fig:reweighting_flowchart} }
\end{figure}

We reweight the prior with the data in 
Table~\ref{eq:input_datasets2} either sequentially (by adding one dataset 
after the other) or simultaneously (by adding all the datasets at once).
In the first case, we monitor the efficiency loss and quantify 
the constraining power of each dataset.
In the second case, we validate the goodness of reweighting against the 
results of a new fit to the extended datasets, by checking that they lead to 
equivalent results (within statistical fluctuations).
Our strategy is schematically summarised in 
Fig.~\ref{fig:reweighting_flowchart}.

\subsection{Monitoring the efficiency loss: the effective number of replicas}

\begin{figure}[!t]
\begin{center}
\includegraphics[width=0.85\linewidth]{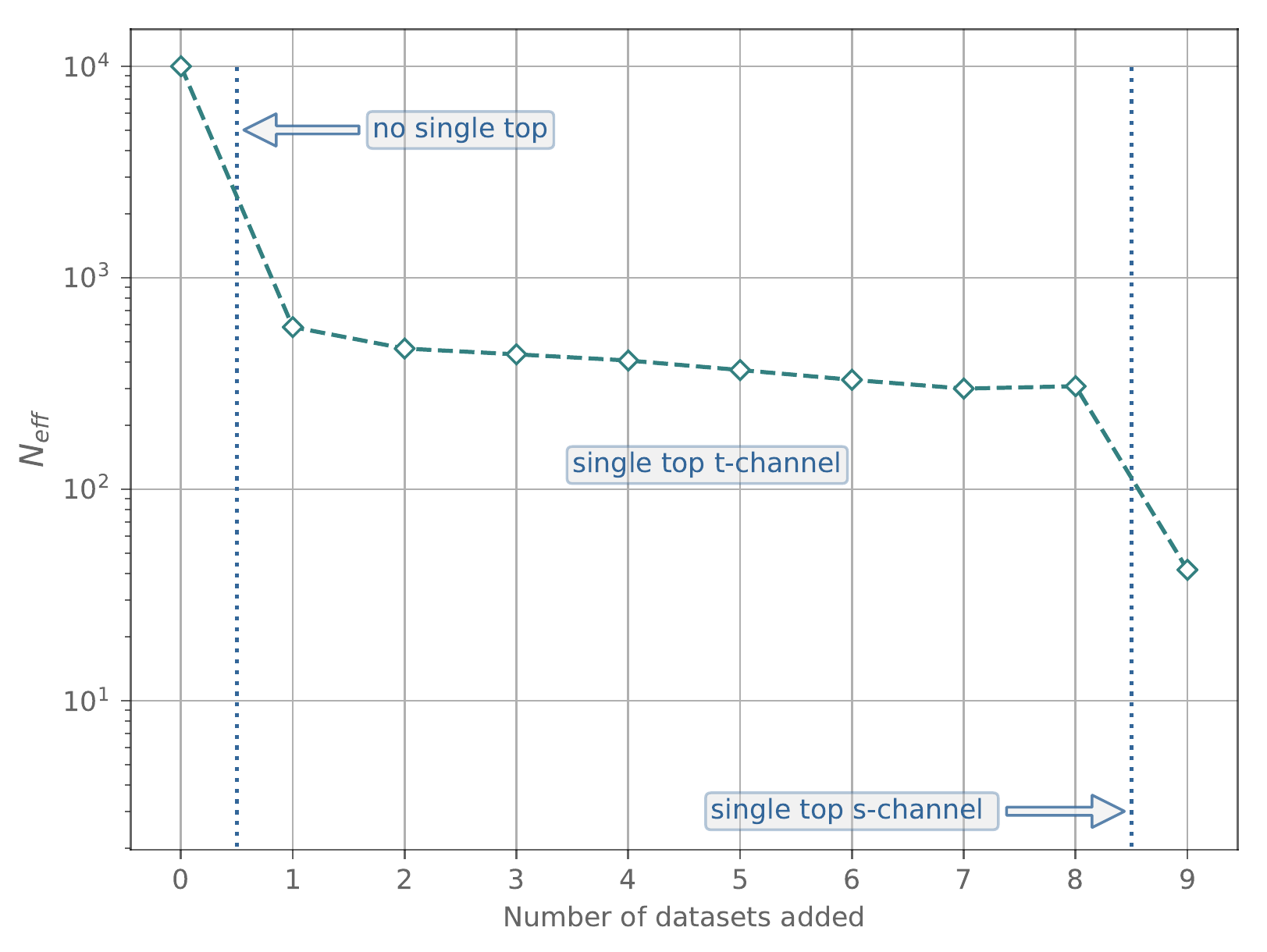}\\
\caption{\small The value of the effective number of replicas $N_{\rm eff}$,
      Eq.~(\ref{eq:neff}), in the prior, which does not contain any
      of the single-top production measurements listed in 
      Table~\ref{eq:input_datasets2}, and once the various single-top datasets
      are sequentially added by reweighting.
      As indicated in the plot, first we add the $t$-channel datasets
      and then the $s$-channel datasets following the order given in 
      Table~\ref{eq:input_datasets2}.
      \label{fig:Neff_singletop} }
  \end{center}
\end{figure}

We first reweight our prior by including sequentially one dataset after the
other, following the order given in Table~\ref{eq:input_datasets2}.
In Fig.~\ref{fig:Neff_singletop} we show the value of the effective number of
replicas $N_{\rm eff}$, Eq.~(\ref{eq:neff}), for each step: point ``0''
corresponds to the prior, which does not contain any
of the single-top production measurements listed in
Table~\ref{eq:input_datasets2}; points ``1''-''8'' correspond to the sets 
reweighted with each of the single-top t-channel datasets;
and point ``9'' corresponds to the set further reweighted with the total 
single-top $s$-channel production cross section from ATLAS at 8 TeV.
Reweighting with the remaining total single-top $s$-channel production cross 
section from CMS at 8 TeV would in principle correspond to an extra point
on the right of Fig.~\ref{fig:Neff_singletop}.
However it is not displayed because the efficiency loss of the reweighted 
ensemble is already significant ($N_{\rm eff}\lesssim 100$) for point ``9''. 
Therefore the corresponding results cannot be trusted.

From Fig.~\ref{fig:Neff_singletop} one observes that the original 
number of replicas in the prior, $N_{\rm rep}=10^4$, are reduced to 
$N_{\rm eff} \simeq 550$ effective replicas once the first single-top 
$t$-channel production dataset is added.
The subsequent addition of the remaining $t$-channel measurements
leads to a further, but mild, decrease of the value of $N_{\rm eff}$
down to around $N_{\rm eff} \simeq 300$. 
This behaviour can be understood if we consider that, the first
time one adds a single-top $t$-channel dataset, one is constraining
several directions in the parameter space that had large uncertainties or
were degenerate in the prior.
Adding subsequent measurements of the same type only refines
the constraints provided by this first dataset.

One may wonder whether the initial abrupt decrease in the effective number of
replicas is just a consequence of the fact that the specific dataset is
inconsistent with the prior.
Computing the $\mathcal{P}(\alpha)$ distribution rules
out this possibility as expected: we know from our previous 
work~\cite{Hartland:2019bjb} that all the datasets in 
Table~\ref{eq:input_datasets2} are consistent with the prior.
For this reason we will refrain from showing the 
$\mathcal{P}(\alpha)$ distribution in the sequel.

Our understanding is further confirmed by observing that
once the $s$-channel measurements are subsequently
added, then $N_{\rm eff}$ falls from $\simeq 300$ to below 50.
Again, there is a large amount of information being added into the probability 
distribution once a completely new type of process is added, since now one 
becomes sensitive to new combinations of SMEFT parameters that are 
unconstrained by the measurements previously considered.
Given that $N_{\rm eff}\simeq 50$, the reliability of the reweighting method 
in this case would be questionable.
Including both the $t$- and the $s$-channel measurements by 
reweighting would require a prior based on a much larger number of replicas, 
{\it e.g.} $N_{\rm rep}=\mathcal{O}(10^5)$.

Finally we have also verified that the order in which specific datasets
are being added via reweighting does not modify the pattern observed in 
Fig.~\ref{fig:Neff_singletop} nor the final result of the procedure.
This behaviour is consistent with what was found
in the PDF case~\cite{Ball:2010gb,Ball:2011gg}.

\subsection{Validation of reweighting: single-top $t$-channel data}

We now reweight our prior by including simultaneously all the single-top
$t$-channel datasets at once.
For the time being, we do not consider single-top $s$-channel datasets, 
because this will result in a too large efficiency loss (see above).
Our aim is to validate the reweighting procedure by comparing the resulting
probability distribution with that obtained from a new fit to the same datasets.

\begin{figure}[!t]
\centering
\includegraphics[width=0.99\linewidth]{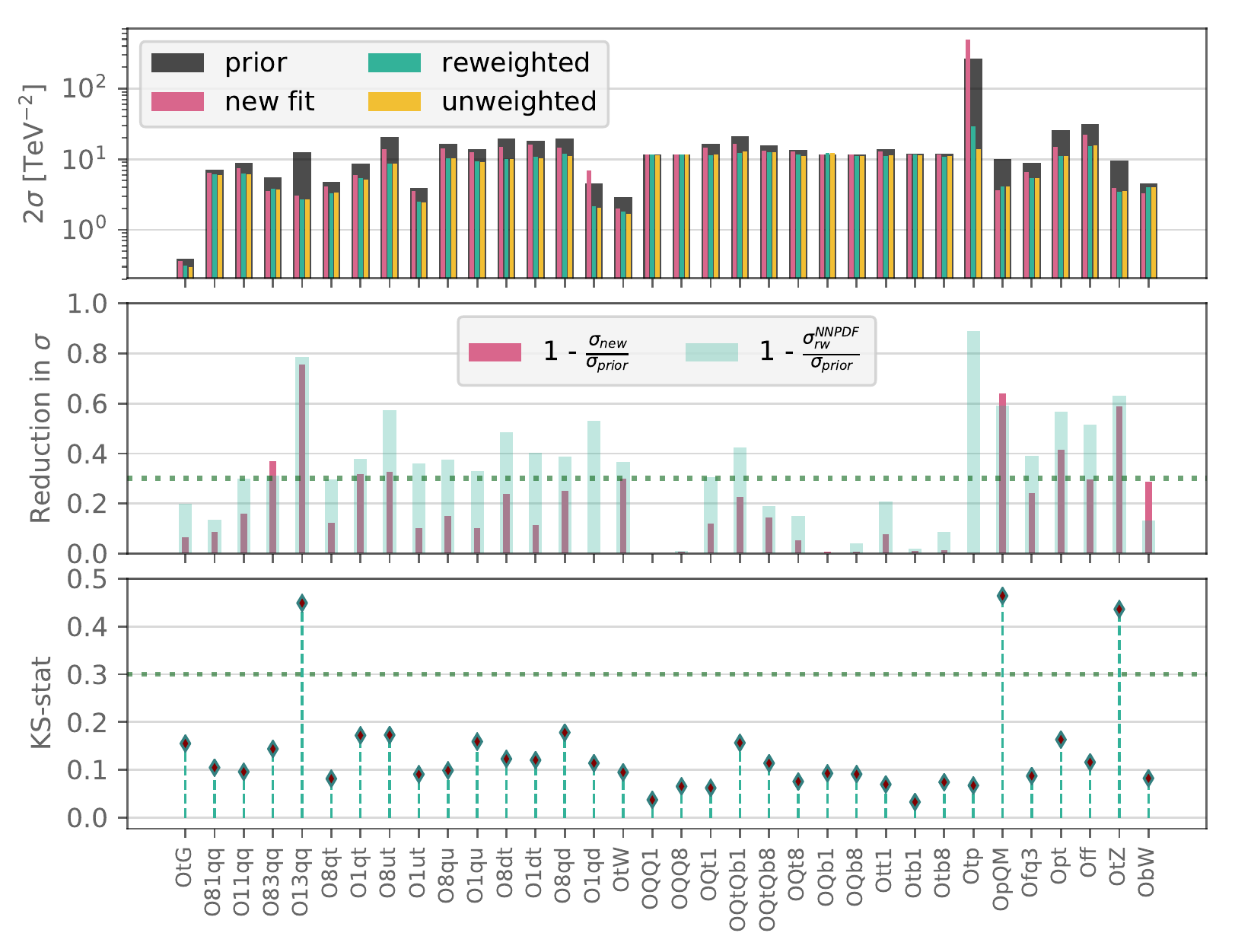}\\
\caption{\small Upper panel: the 95\% CL bounds for
    the $N_{\rm op}=34$ Wilson coefficients considered in this SMEFT analysis 
    of the top quark sector.
    We compare the prior results (without any $t$- or $s$-channel single-top 
    production data included) with those after the $t$-channel measurements 
    have been added either by reweighting or by performing a new fit.   
    Central panel: the relative 68\% CL uncertainty reduction compared to 
    the prior, both for the reweighted and for the new fit cases.
    Lower panel: the associated value of the KS statistic
    computed between the unweighted and the prior results.
    In both the central and lower panels, the horizontal dotted lines 
    indicate the thresholds to select the operators for which 
    Bayesian reweighting is meaningful.
\label{fig:two_sigma_bounds_t_channel} }
\end{figure}

In the upper panel of Fig.~\ref{fig:two_sigma_bounds_t_channel} we compare the 
results obtained from reweighting and from the new fit.
Specifically we show the 95\% CL bounds for the $N_{\rm op}=34$ Wilson 
coefficients considered in our previous SMEFT analysis of the top quark 
sector~\cite{Hartland:2019bjb}.
Note that we are assuming that $\Lambda=1$ TeV.
For completeness, we also show the corresponding unweighted results: 
in all cases we find excellent agreement with the reweighted results. 
We will thus treat them as equivalent in the following.

From this comparison one finds that the results obtained from reweighting or
from a new fit are reasonably similar in most cases.
To facilitate their interpretation, we compare the relative 
68\% CL uncertainty reduction between the reweighted and the prior cases,
$1-\sigma_{\rm rw}^{\rm NNPDF}/\sigma_{\rm prior}$, and between the 
new fit and the prior cases, $1-\sigma_{\rm new}/\sigma_{\rm prior}$.
We observe that for three degrees of freedom the reweighting of the prior 
with the $t$-channel single-top cross section data leads to a 
reduction of the uncertainties larger than a factor of two,
consistently with the new fit.
These are the Wilson coefficients associated to the {\tt O13qq}, {\tt OpQM} 
and {\tt OtZ} SMEFT operators (we will henceforth use the notation 
of~\cite{Hartland:2019bjb}).
Not surprisingly these are the three operators for which adding $t$-channel 
single-top data to the prior has the largest effect.
In particular, {\tt O13qq} and {\tt OpQM} are either directly (or indirectly,
via correlations with other coefficients) constrained by $t$-channel single-top 
data.
The reason {\tt OtZ} is also more constrained is because the data either 
provides access to a previously unconstrained direction in the SMEFT parameter 
space, see Table 3.5 in~\cite{Hartland:2019bjb},
or because it breaks degeneracies between directions.
We therefore conclude that reweighting leads to results equivalent to those of 
a new fit for the operators that are being more directly constrained by 
the new data.

If we now look at other operators, we still clearly find that reweighting
leads to a reduction of the 95\% CL bounds in comparison to the prior.
However such a reduction seems sometimes over-optimistic, especially
if it is compared to the new fit results (see for example the {\tt Otp} 
or {\tt O1qd} operators).
In this case, reweighting seemingly fails.
Nevertheless the 95\% CL bounds are only a rough measure of the 
actual change in the probability distributions from the prior to the 
reweighted ensemble.
Their interpretation is particularly unclear when statistical 
fluctuations (including from finite-size effects intrinsic 
to a Monte Carlo analysis such as the current one) become large.
This mostly happens for poorly-constrained operators.

To discriminate whether the discrepancies observed in 
Fig.~\ref{fig:two_sigma_bounds_t_channel} are induced by
a genuine change in the probability distributions of the various
operators or by a statistical fluctuation, we look at the
corresponding KS statistic, Eq.~\eqref{eq:KS-test}.
Only when the value of the KS statistic is sufficiently large,
can one claim that the differences between the prior and the reweighted
distributions are statistically significant.
With this motivation in mind, in the lower panel of
Fig.~\ref{fig:two_sigma_bounds_t_channel} we display, for each Wilson 
coefficient, the values of the KS statistic computed
between the unweighted and the prior distributions.
The largest values of the KS statistic are associated to 
the {\tt O13qq}, {\tt OpQM} and {\tt OtZ} operators.
These are precisely the operators for which we know that the data has the 
largest effect and for which reweighting is equivalent to a new fit.
Low values of the KS statistic are associated to most of the other 
operators, including those for which the reduction of the 95\% CL bounds
induced by reweighting is seemingly large (and even much larger than 
the reduction induced by a new fit). 
This is the case, {\it e.g.}, for the {\tt O1qd} and {\tt Otp} operators. 

The results of Fig.~\ref{fig:two_sigma_bounds_t_channel} 
show that, in a global SMEFT analysis, reweighting successfully reproduces 
the results of a new fit when the two following conditions are satisfied:
\begin{itemize}

\item the size of 95\% CL intervals of a specific operator after 
  reweighting is reduced by an amount higher than a given threshold;

\item the value of the KS statistic is sufficiently high to ensure that 
  the modification in the probability distributions is not induced by
  a statistical fluctuation.
  
\end{itemize}
  
Of course there will always be some ambiguity when setting the thresholds
for the 95\% CL bound reductions and for the KS statistic.
In Fig.~\ref{fig:two_sigma_bounds_t_channel} we indicate two possible values of 
these thresholds in the central and lower panels with dotted horizontal lines: 
$1-\sigma_{\rm rw}^{\rm NNPDF}/\sigma_{\rm prior}=1-\sigma_{\rm new}/\sigma_{\rm prior}=0.3$ and ${\rm KS}=0.3$.
These values will select {\tt O13qq}, {\tt OpQM} and {\tt OtZ} as the only 
three out of the $N_{\rm op}=34$ operators for which the reweighted results 
are reliable.
For the remaining operators, the two conditions above will not be
satisfied and the corresponding reweighting results could not be trusted.
While the selection criteria adopted here are mostly
based on phenomenological evidence, it would be advantageous
to derive more formal criteria that could be used in general
situations. 
We defer such an investigation to future work.

\begin{figure}[!t]
\centering
\includegraphics[width=0.49\textwidth]{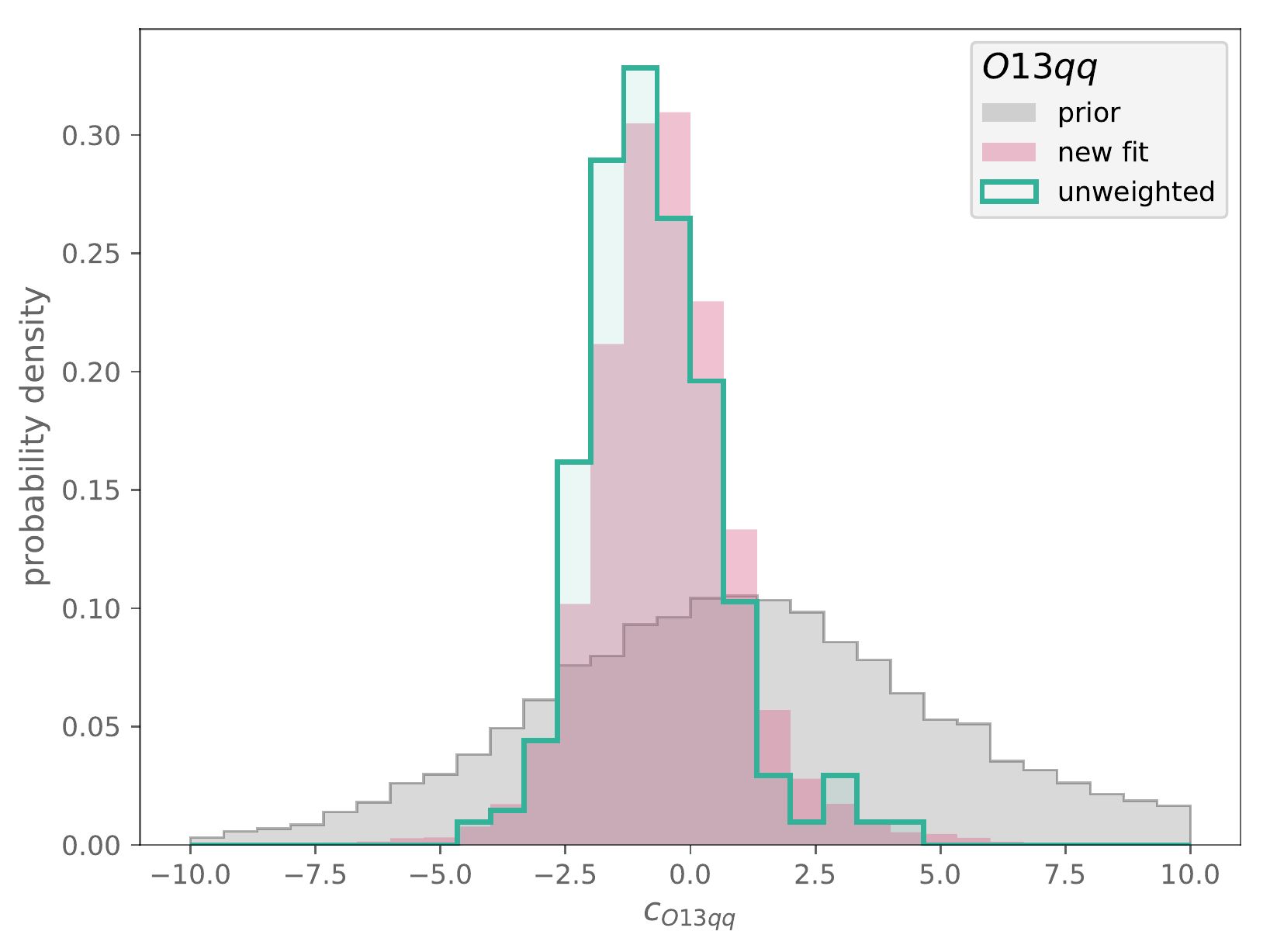}
\includegraphics[width=0.49\textwidth]{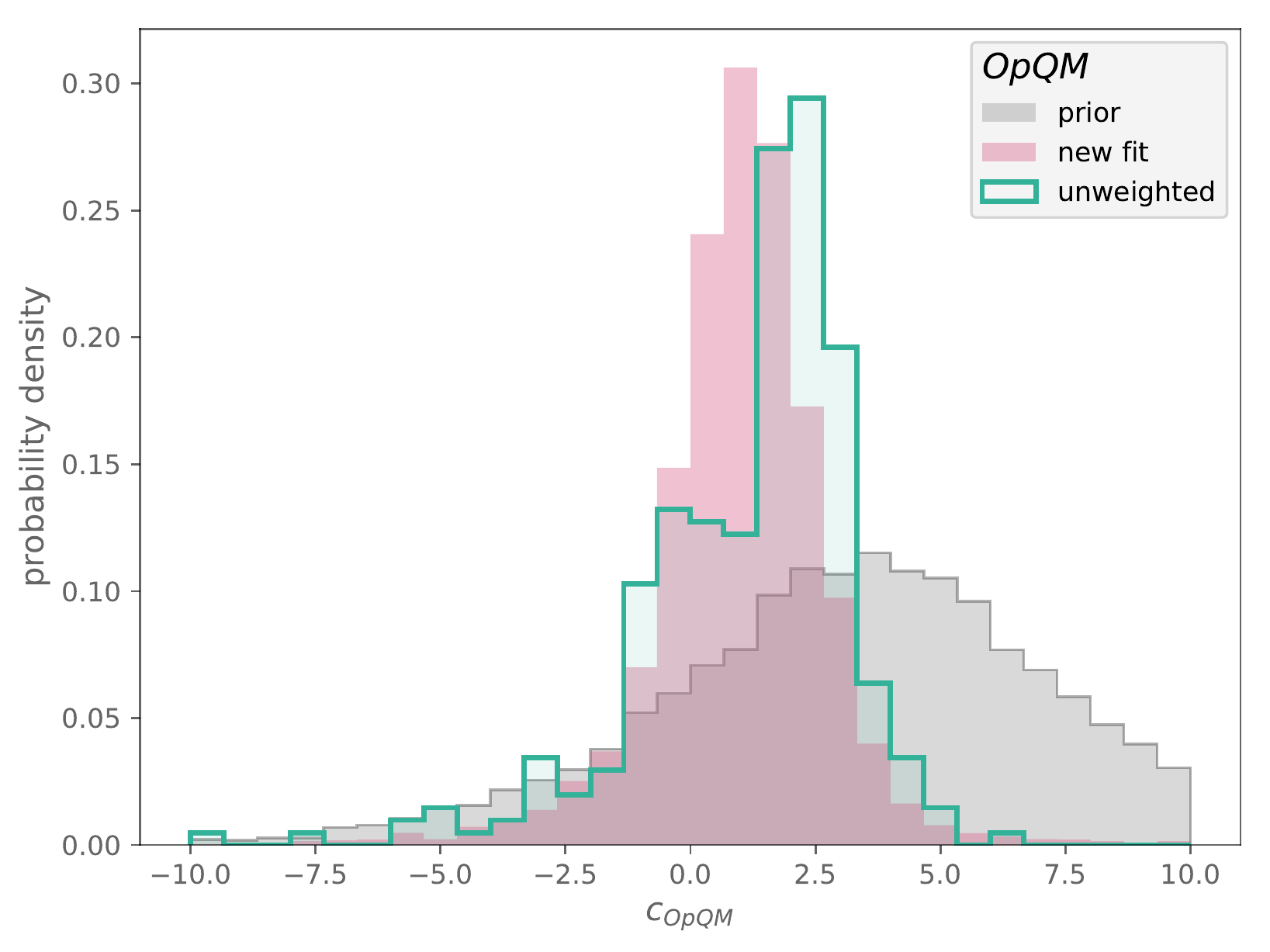}\\
\includegraphics[width=0.49\textwidth]{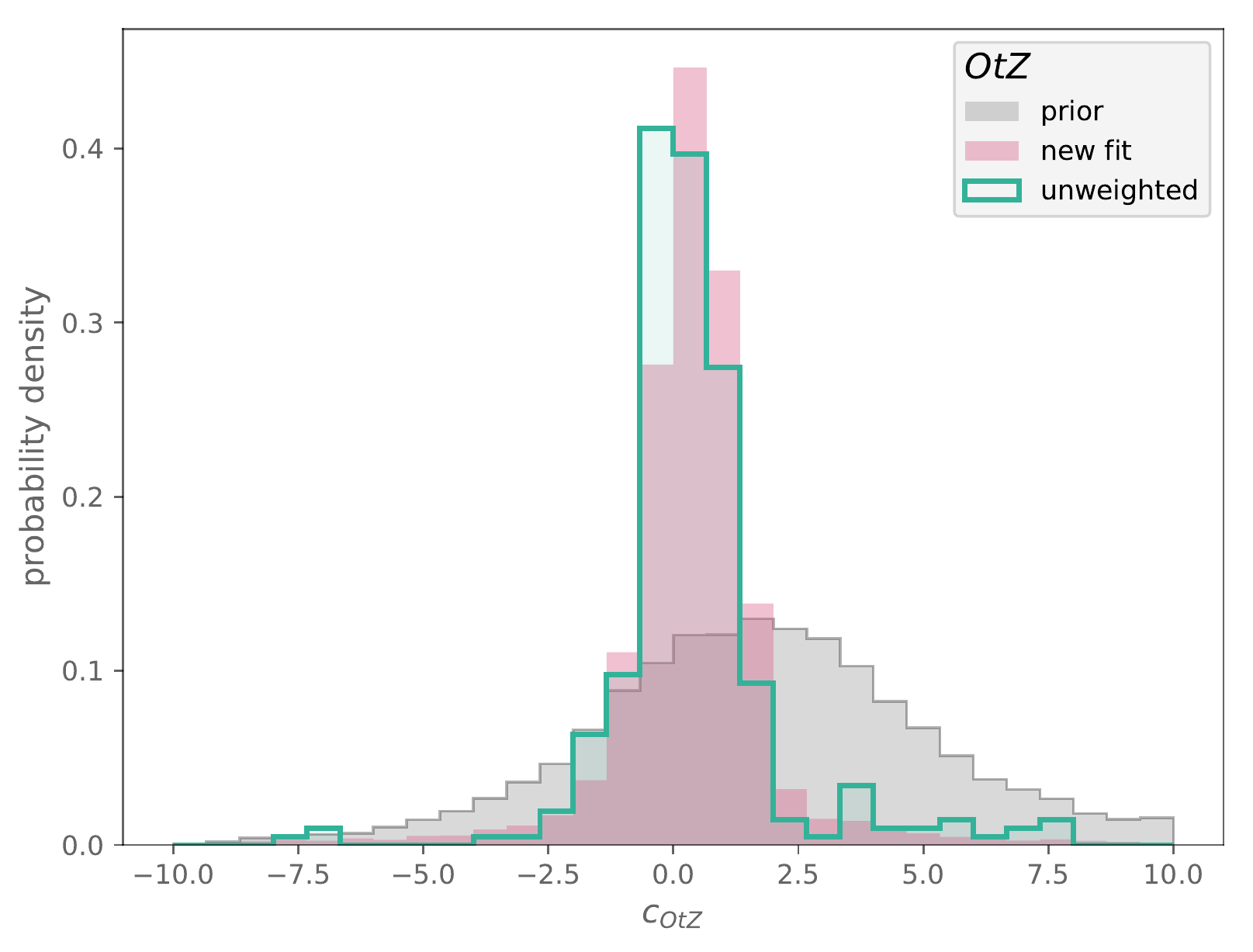}\\
\caption{\small The probability distribution associated to the Wilson 
      coefficients $c_{qq}^{13}, c_{pQM}$, and $c_{tZ}$. The prior results
      are compared to the results obtained by reweighting and unweighting the
      prior with the $t$-channel single-top production cross sections and 
      from a new fit to the extended datasets.
      \label{fig:t_ch_coeff_dists}  }
\end{figure}

As a final cross-check, the probability distributions of the three operators 
associated to the Wilson coefficients for which reweighting is applicable, 
{\tt O13qq}, {\tt OpQM} and {\tt OtZ}, are displayed 
in Fig.~\ref{fig:t_ch_coeff_dists}.
The prior results are compared to the results obtained by reweighting and 
unweighting the prior with the $t$-channel single-top production cross section 
data and the results obtained from a new fit to the same set of data.
The prior and the new fit sets are made of $N_{\rm rep}=10^4$ replicas; the
unweighted set is made from $N_{\rm eff}= 300$ effective replicas.
First, we observe how the prior distribution is significantly narrowed once the 
new data is added, either by reweighting or by a new fit, consistently with the 
results displayed in the central panel of 
Fig.~\ref{fig:two_sigma_bounds_t_channel}.
Second, we observe good agreement between the reweighted and the new fit
shapes of the probability distributions, despite the former being based on a 
much smaller number of replicas than the latter.
All this is consistent with the high value of the KS statistic 
associated to the three operators under examination.

In this work we have so far validated the reweighting approach using theory
calculations that included $\mathcal{O}\lp \Lambda^{-4}\rp$
corrections in the prior fit, in the reweighted results and in the new fit.
However, the validity of Bayesian reweighting should be
independent of the specific theory assumptions used.
In this respect, we have verified that an agreement between
reweighting and the new fit similar to that reported
in Fig.~\ref{fig:two_sigma_bounds_t_channel}
is obtained  when only $\mathcal{O}\lp \Lambda^{-2}\rp$
corrections are included in the SMEFT theory calculation.
This exercise demonstrates that Bayesian reweighting
can be applied in the same way in both cases.
It is up to the users to decide on their preferred theory settings ---
our {\tt SMEFiT} results are available both at
$\mathcal{O}\lp \Lambda^{-2}\rp$ and at $\mathcal{O}\lp \Lambda^{-4}\rp$.
  
Note that, in general, the subset of operators which are more affected by the 
new data differs between the $\mathcal{O}\lp \Lambda^{-2}\rp$ and the 
$\mathcal{O}\lp \Lambda^{-4}\rp$ fits.
In the case of the $t$-channel single top production
measurements, the three operators for which the impact
of the new data is more marked are: {\tt O13qq}, {\tt OpQM}, and {\tt OtZ}
at  $\mathcal{O}\lp \Lambda^{-4}\rp$; {\tt O13qq}, {\tt OpQM}, {\tt OtW} and 
{\tt Opt} at $\mathcal{O}\lp \Lambda^{-2}\rp$.
The fact that the specific operators which are more constrained by a new piece 
of experimental information depend on whether the calculation is performed at
$\mathcal{O}\lp \Lambda^{-2}\rp$ or the $\mathcal{O}\lp \Lambda^{-4}\rp$
is expected since in general each calculation probes different regions
of the parameter space, as discussed in~\cite{Hartland:2019bjb}.

A final remark concerns the interpretation
of the results presented in Fig.~\ref{fig:two_sigma_bounds_t_channel}.
Within a SMEFT analysis one is only sensitive to the ratio $c_k/\Lambda^2$ 
rather than to the absolute New Physics scale $\Lambda$.
While here we assume $\Lambda=1$ TeV for illustrative purposes, it
is possible to interpret our results for any other value of $\Lambda$.
In particular, the upper (lower) bounds on the $k$-th Wilson coefficient, 
$\delta c_k^{+}$ ($\delta c_k^{-}$), should be rescaled as
\be
  \label{eq:rescalingLambda}
  \delta\widetilde{c}^{\pm}_k = \delta c_k^{\pm} \times \lp \frac{\widetilde{\Lambda}}{\Lambda}\rp^2 \,
\ee
in comparison to the results shown here for the case 
$\widetilde{\Lambda} \ne \Lambda = 1$ TeV.
This said, the validity of Bayesian reweighting is
independent from the interpretation of the results in terms of a specific
value of $\Lambda$.
The important discussion about which values of $\Lambda$ can be probed
when interpreting the results to ensure the validity of the EFT regime, see
for instance~\cite{Hartland:2019bjb} and references therein, should thus be 
separated from the validation of Bayesian reweighting.
  

\subsection{Independence from the process type: single-top $s$-channel data}

We now repeat the exercise carried out in the previous subsection by 
simultaneously reweighting our prior with all the single-top $s$-channel data
listed in Table~\ref{eq:input_datasets2}, {\it i.e.} the two cross sections
from ATLAS and CMS at 8 TeV.
Despite having only $n_{\rm dat}=2$ additional data points, one can in principle
expect to improve the prior by a significant amount because the new process
probes different top quark couplings with respect to those already included
in the prior.
By doing so, our purpose is to check whether the conclusions reached in the 
case of single-top $t$-channel datasets can be extended to datasets for 
processes of a different type.

\begin{figure}[!t]	
\centering
\includegraphics[width=0.99\linewidth]{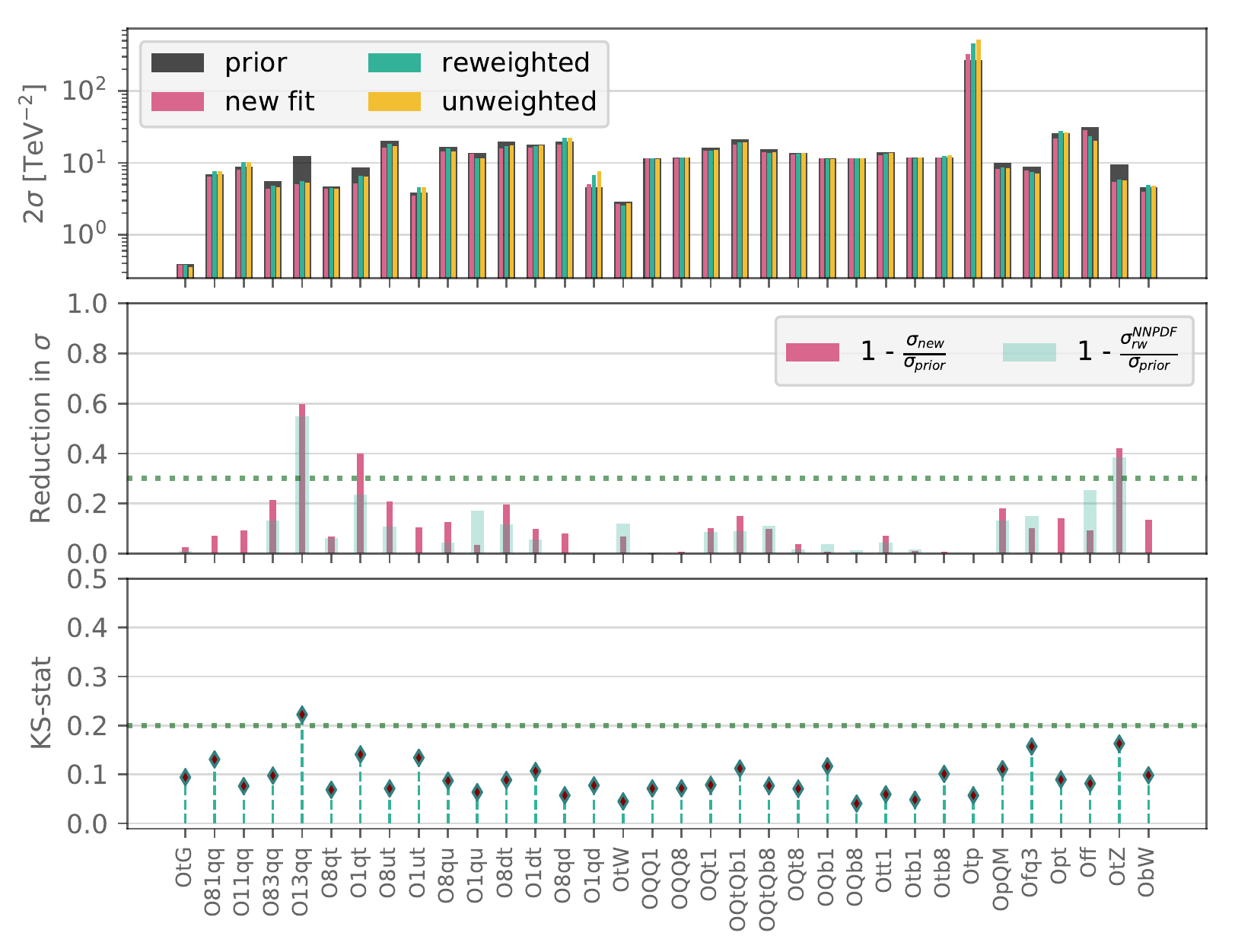}\\
\caption{\small Same as Fig.~\ref{fig:two_sigma_bounds_t_channel},
          but in the case of single-top $s$-channel production total 
          cross sections at 8 TeV.
          The selection threshold for the KS statistic is set to 0.2
          in this case.
\label{fig:two_sigma_bounds_s_channel}	}
\end{figure}

In Fig.~\ref{fig:two_sigma_bounds_s_channel} we display the same results as in
Fig.~\ref{fig:two_sigma_bounds_t_channel}, but for the case of $s$-channel 
single-top production total cross sections.
The impact of the $n_{\rm dat}=2$ $s$-channel data points is rather smaller than 
the impact of the $n_{\rm dat}=25$ $t$-channel ones, though still appreciable.
The values of the KS statistic are consequently small for all the 
$N_{\rm op}=34$ operators but one: {\tt O13qq}.
In this case, this operator is the only one that satisfies the selection 
criteria defined above, whereby ${\rm KS} \ge 0.2$ and 
$1-\sigma_{\rm rw}^{\rm NNPDF}/\sigma_{\rm prior}= 1-\sigma_{\rm new}/\sigma_{\rm prior}\ge 0.3$.
As expected, good agreement is found between the reweighted and the new fit 
results for the {\tt O13qq} operator.

As discussed above there is an irreducible ambiguity in the choice of the 
threshold values for the relative reduction of the 95\% CL bounds and of the 
KS statistic.
For instance, if we look at the {\tt OtZ} operator, the inclusion of the
single-top $s$-channel cross sections leads to an uncertainty reduction of 
around 40\% and to a KS statistic of ${\rm KS} \simeq 0.18$; 
reasonable agreement between reweighted and new fit results is also observed 
for {\tt OtZ}.
However, this operator does not pass validation if we require ${\rm KS}\ge 0.3$.
Therefore it is up to the user to decide how conservative he wants to be:
the higher the selection thresholds, the more reliable the reweighting 
results will be.

\begin{figure}[!t]
\centering
\includegraphics[width=0.49\textwidth]{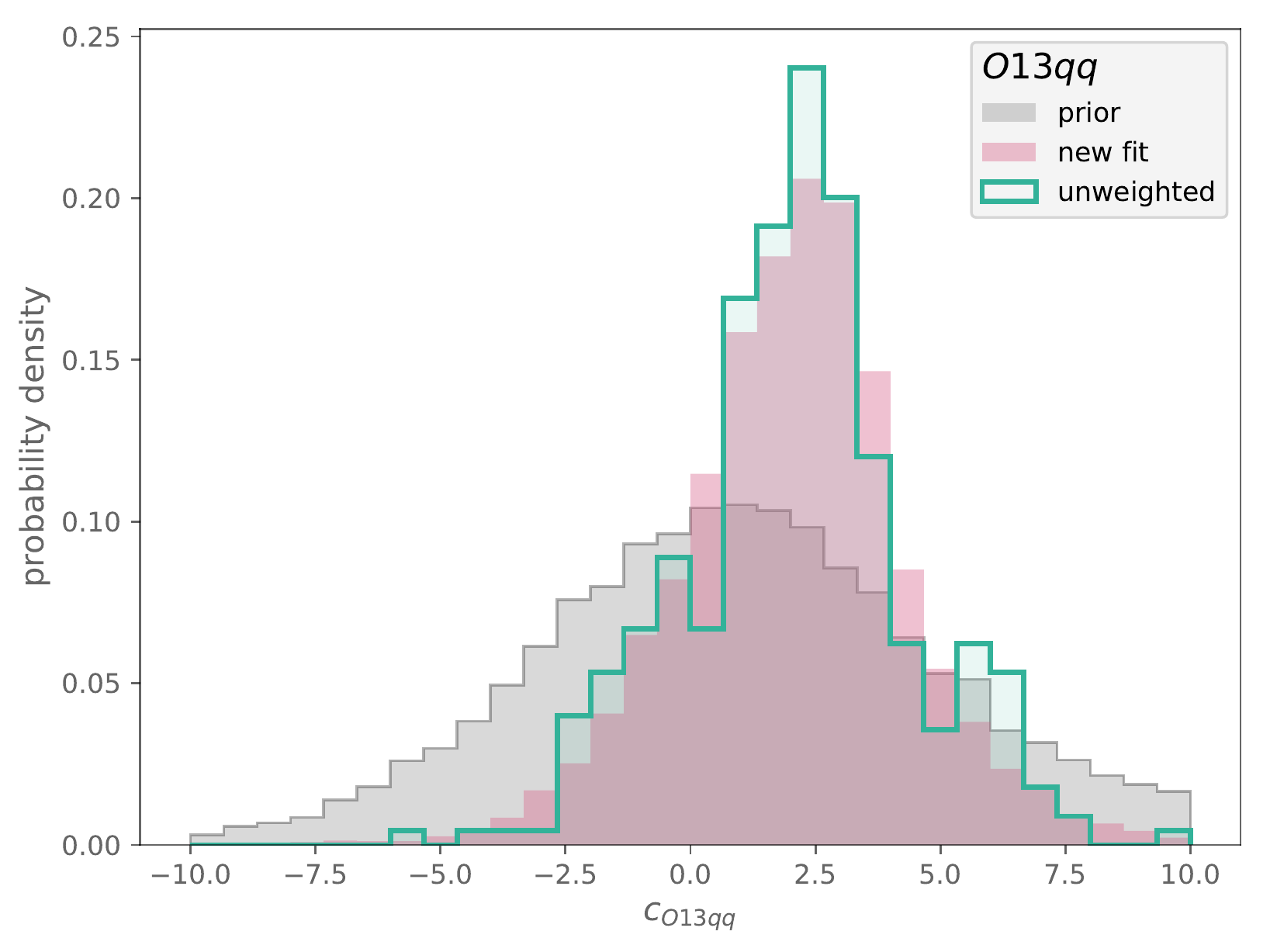}
\includegraphics[width=0.49\textwidth]{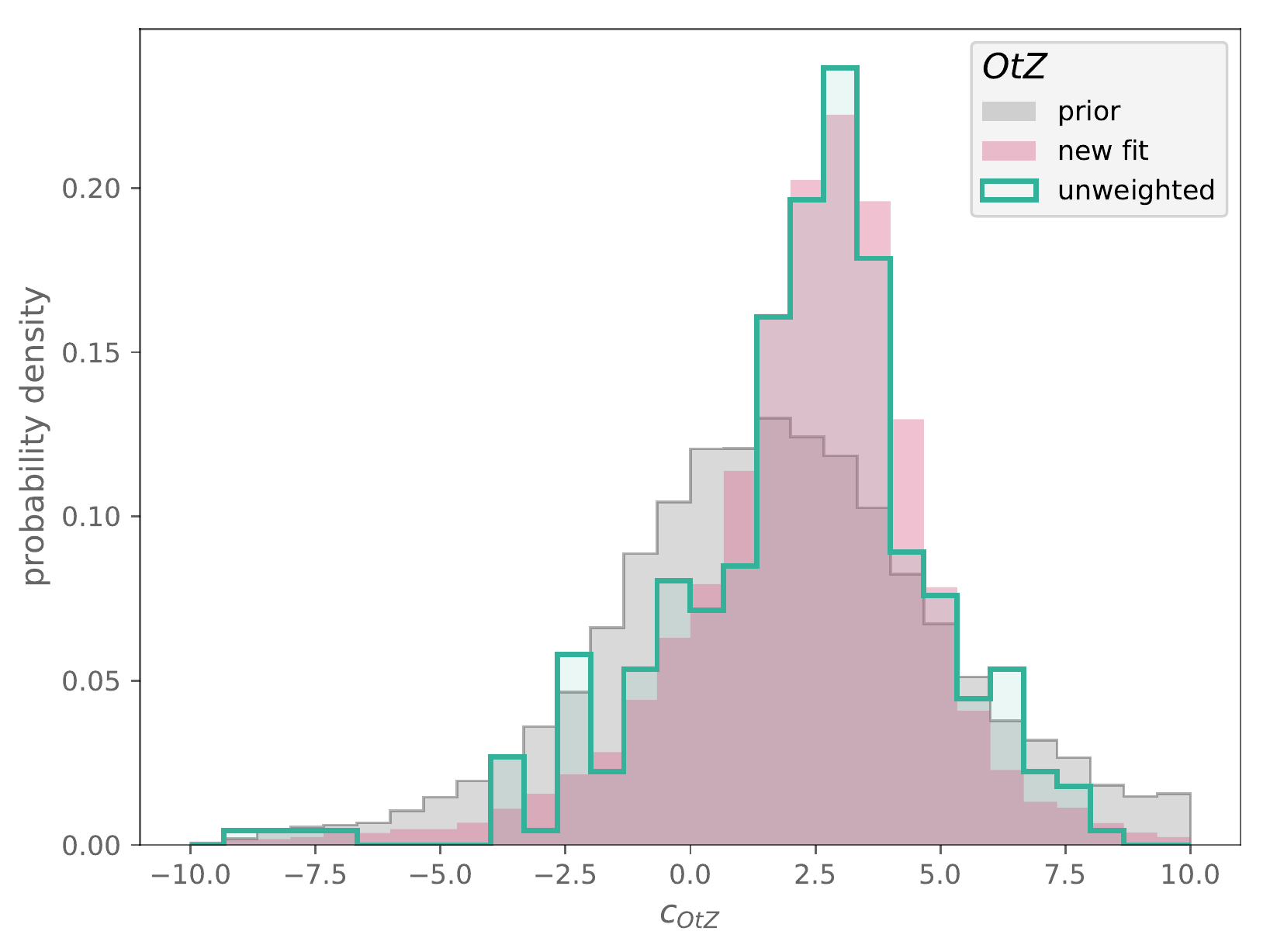}\\
\caption{\small Same as  Fig.~\ref{fig:t_ch_coeff_dists} but for the 
         Wilson coefficients mostly constrained by single-top $s$-channel 
         production cross sections.
	\label{fig:s_ch_coeff_dist}   }
\end{figure}

Finally in Fig.~\ref{fig:s_ch_coeff_dist} we repeat the comparison shown in
Fig.~\ref{fig:t_ch_coeff_dists}, but for the probability distributions of the 
Wilson coefficients associated to the {\tt O13qq} and {\tt OtZ} SMEFT operators.
We show the results obtained from the prior, from reweighting and unweighting 
it with the single-top $s$-channel production cross sections and from a new
fit to the same dataset.
We find good agreement between the unweighted and the new fit results and 
observe how the relative narrowing of the distribution is less marked than 
in the case of single-top $t$-channel cross sections.
This is understood as single-top $s$-channel measurements have less constraining
power than single-top $t$-channel measurements once included in the prior.

%% file: tables/input-datasets.tex
\begin{table}[!t]
\centering
\small
\renewcommand{\arraystretch}{1.70}
\begin{tabular}{clccccc}
\toprule
ID           &
Dataset      &  
$\sqrt{s}$   & 
Info         &  
Observables  & 
$n_{\rm dat}$  &  
Ref.         \\
\midrule
1                                             &
{\tt CMS\_t\_tch\_8TeV\_dif}                  &
{\bf 8 TeV}                                   &  
$t$-channel                                   & 
$d\sigma/d|y^{(t+\bar{t})}|$                     & 
6                                             & 
\cite{CMS-PAS-TOP-14-004}                     \\
2                                             &
{\tt ATLAS\_t\_tch\_8TeV}                     &
{\bf 8 TeV}                                   & 
$t$-channel                                   &
$d\sigma(t)/dy_t$                             &
4                                             & 
\cite{Aaboud:2017pdi}                         \\
3                                             &
{\tt ATLAS\_t\_tch\_8TeV}                     &
{\bf 8 TeV}                                   & 
$t$-channel                                   &
$d\sigma(\bar{t})/dy_{\bar{t}}$                 &
4                                             & 
\cite{Aaboud:2017pdi}                         \\
4                                             &
{\tt CMS\_t\_tch\_13TeV\_dif}                 &
{\bf 13 TeV}                                  &  
$t$-channel                                   & 
$d\sigma/d|y^{(t+\bar{t})}|$                     & 
4                                             & 
\cite{CMS:2016xnv}                            \\
5                                             &
{\tt CMS\_t\_tch\_8TeV\_inc}                  &
{\bf 8 TeV}                                   & 
$t$-channel                                   &
$\sigma_{\rm tot}(t),\sigma_{\rm tot}(\bar{t})$   & 
2                                             &
\cite{Khachatryan:2014iya}                    \\
6                                             &
{\tt CMS\_t\_tch\_13TeV\_inc}                 &
{\bf 13 TeV}                                  & 
$t$-channel                                   &
$\sigma_{\rm tot}(t+\bar{t})$                   & 
1                                             &
\cite{Sirunyan:2016cdg}                       \\
7                                             &
{\tt ATLAS\_t\_tch\_8TeV}                     &
{\bf 8 TeV}                                   & 
$t$-channel                                   &
$\sigma_{\rm tot}(t),\sigma_{\rm tot}(\bar{t})$   & 
2                                             &
\cite{Aaboud:2017pdi}                         \\
8                                             &
{\tt ATLAS\_t\_tch\_13TeV}                    &
{\bf 13 TeV}                                  & 
$t$-channel                                   &
$\sigma_{\rm tot}(t),\sigma_{\rm tot}(\bar{t})$   & 
2                                             &
\cite{Aaboud:2016ymp}                         \\
\midrule
9                                             &
{\tt ATLAS\_t\_sch\_8TeV}                     &
{\bf 8 TeV}                                   & 
$s$-channel                                   &
$\sigma_{\rm tot}(t+\bar{t})$                   & 
1                                             &
\cite{Aad:2015upn}                            \\
10                                            &
{\tt CMS\_t\_sch\_8TeV}                       &
{\bf 8 TeV}                                   & 
$s$-channel                                   &
$\sigma_{\rm tot}(t+\bar{t})$                   & 
1                                             &
\cite{Khachatryan:2016ewo}                    \\
\bottomrule
  \end{tabular}
  \caption{\small The measurements of single-top quark production 
    at the LHC (both in the $t$-channel and in the $s$-channel) used in this 
    analysis to validate the results of Bayesian reweighting.
    For each dataset, we indicate the dataset label, the center of mass energy 
    $\sqrt{s}$, the production mechanism, the type of observables, 
    the number of data points $n_{\rm dat}$, and the publication reference.
    \label{eq:input_datasets2}
  }
\end{table}

%% file: sec-gk.tex
\section{Dependence on the choice of weights}
\label{sec:sec-gk}

The results presented in the previous section are based on the expression for 
the weights given by Eq.~\eqref{eq:weightsRW}.
This formula was originally derived in~\cite{Ball:2010gb,Ball:2011gg} and
has been routinely used to quantify the impact of new data in studies of PDFs 
since then.
Results obtained with Eq.~\eqref{eq:weightsRW} were found to be equivalent to 
the results obtained with a new fit of the data in all cases, and were even 
benchmarked in a closure test~\cite{Ball:2014uwa}.
In this work, we showed that Eq.~\eqref{eq:weightsRW} works 
also in the case of a global SMEFT analysis, provided specific selection 
criteria for the individual operators are satisfied.
In this section, we study the dependence of our results upon the choice of 
weights.

\subsection{Giele-Keller weights}

An expression for the weights different from the one in Eq.~\eqref{eq:weightsRW}
has been suggested in the past.
Specifically, in a formulation which pre-dates the one 
in~\cite{Ball:2010gb,Ball:2011gg}, Giele and Keller 
advocated~\cite{Giele:1998gw,Giele:2001mr} that the weights should read instead
\be
\label{eq:weightsGK}
\omega_k \propto \exp\left( -\chi^2_k/2\right) \, ,\quad k=1,\ldots, N_{\rm rep} \, ,
\ee
where, in comparison to Eq.~\eqref{eq:weightsRW}, the prefactor 
$(\chi_k^2)^{(n_{\rm dat}-1)/2}$ is dropped.
We will refer to Eq.~\eqref{eq:weightsRW} and to Eq.~\eqref{eq:weightsGK}
as NNPDF and GK weights, respectively, in the remainder of this section.
The main difference between NNPDF and GK weights is that, for consistent data, 
the largest weights are assigned respectively to replicas either 
with $\chi^2_k \simeq n_{\rm dat}$  or with $\chi^2_k\to 0$.
If Eq.~\eqref{eq:weightsGK} is used instead of Eq.~\eqref{eq:weightsRW},
a replica associated to $\chi^2_k\to 0$ is not treated as an outlier and 
discarded, as it should, but it is assigned a large weight.

In order to explore the dependence of the reweighted results on the choice of 
the weight formula, we repeat the exercise presented in the previous section 
by using the GK weights instead of the NNPDF weights.
By comparison with our previous results, we expect to determine whether the
GK formula reproduces the results of a new fit as well as the NNPDF
formula, and if it does so more efficiently.
In principle, for $N_{\rm rep}\to \infty$, it is conceivable that the NNPDF and 
the GK formul{\ae} lead to indistinguishable results, which then become 
different only because of finite-size effects.

\begin{figure}[!t]
\centering
\includegraphics[width=0.99\linewidth]{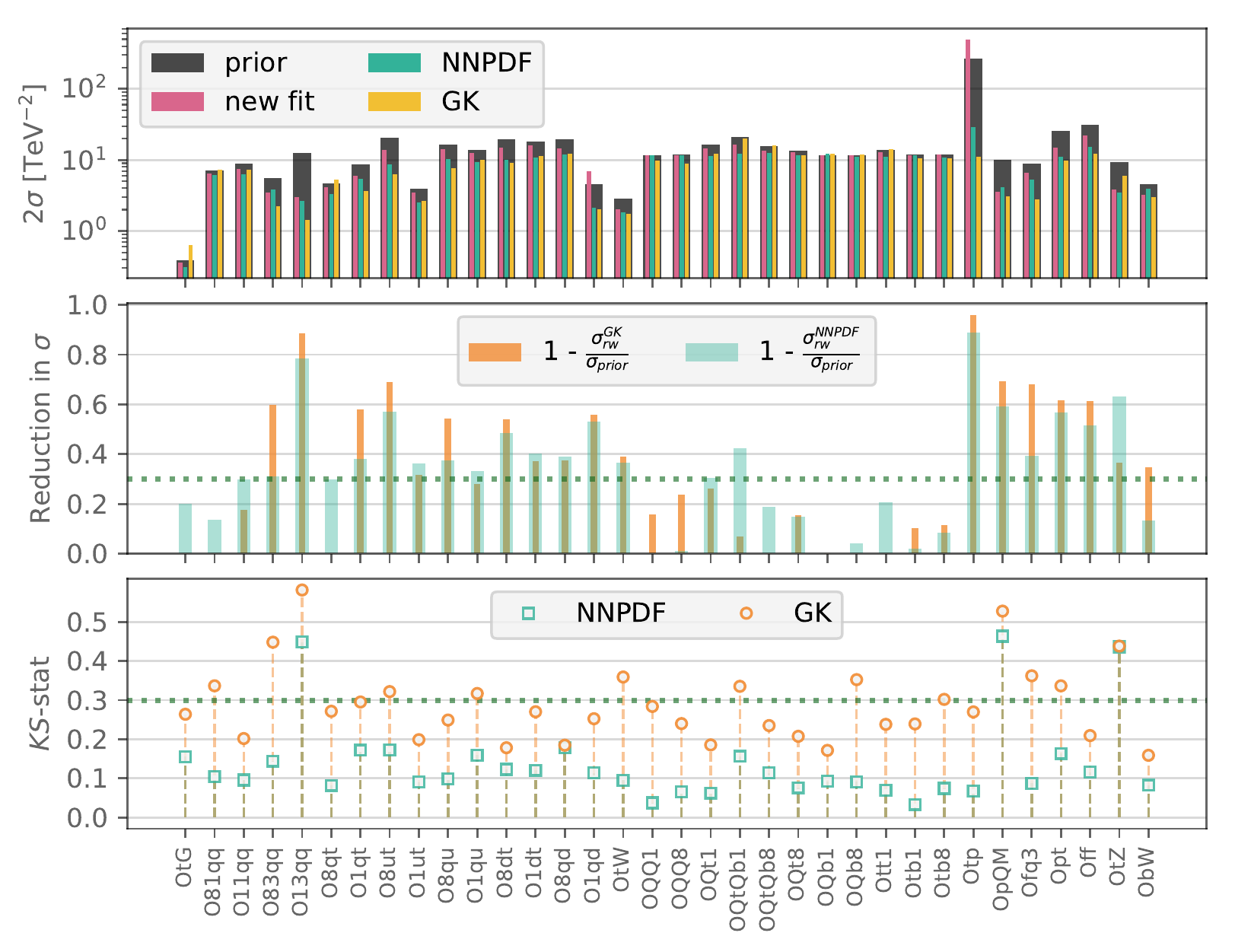}
\caption{\small Same as Fig.~\ref{fig:two_sigma_bounds_t_channel}, but for
            reweighted results obtained either with the NNPDF or the GK weights.
\label{fig:gk_vs_nnpdf_full_t_channel}  }	
\centering
\end{figure}

In Fig.~\ref{fig:gk_vs_nnpdf_full_t_channel} we compare the results obtained
by reweighting the prior with all the single-top $t$-channel data in 
Table~\ref{eq:input_datasets2} either with the NNPDF or the GK weight formula.
The format of the results is the same as in 
Fig.~\ref{fig:two_sigma_bounds_t_channel}, {\it i.e.} each panel displays, 
from top to bottom, the 95\% CL bounds, the corresponding relative reduction 
with respect to the prior and the KS statistic.

We recall that the reweighted results obtained with the NNPDF weights 
reproduce the results obtained with a new fit only for the three operators 
that are more directly constrained by the new data: 
{\tt O13qq}, {\tt OpQM} and {\tt OtZ}.
A meaningful comparison with the results obtained with GK weights
should therefore first focus on these three operators.
By inspection of Fig.~\ref{fig:gk_vs_nnpdf_full_t_channel} such a comparison
reveals that NNPDF and GK results can be rather different.
In particular, the GK results can either overestimate (for {\tt O13qq} and
{\tt OpQM}) or underestimate (for {\tt OtZ}) the uncertainty reduction.

\begin{figure}[!t]
\centering
\includegraphics[scale=0.70]{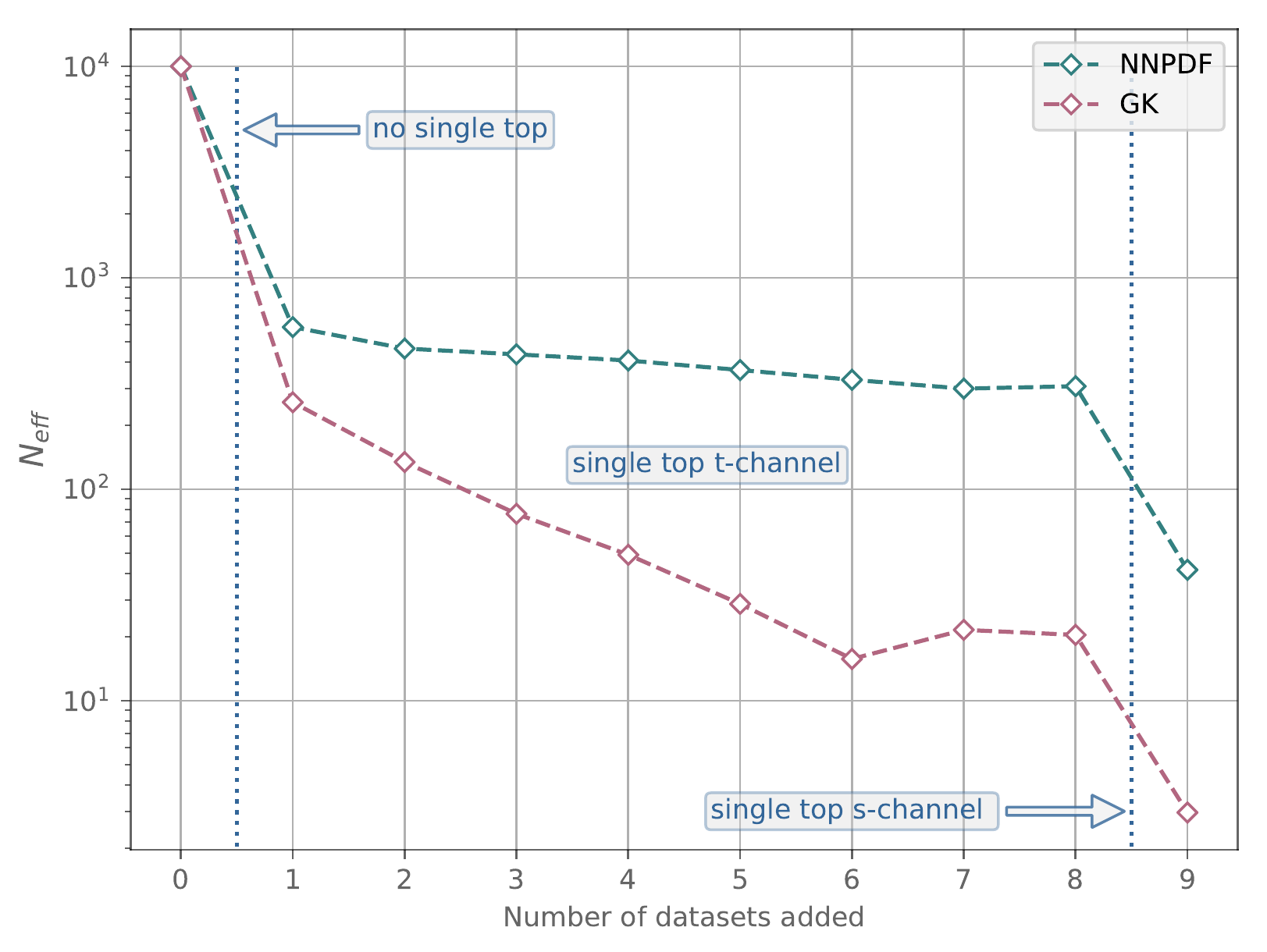}
\caption{\small Same as Fig.~\ref{fig:Neff_singletop}, now
            comparing the effective number of replicas $N_{\rm eff}$
            upon the addition of new data using either the NNPDF or the GK
            weights.
            In both cases, the prior is the same and it is made of 
            $N_{\rm rep}=10^4$ replicas.
	 \label{fig:neff_vs_data}  }	
\end{figure}

Marked differences between NNPDF and GK results persist across all the 
operators.
The values of the KS statistic is consistently larger in the GK case than 
in the NNPDF case.
We therefore investigate the behaviour of the efficiency in the GK case.
In Fig.~\ref{fig:neff_vs_data} we show the dependence of the effective number
of replicas $N_{\rm eff}$ upon the addition of new data both in the 
NNPDF and in the GK cases.
Given that the same amount of new information is added, it is apparent that
the effective number of replicas decreases much faster for GK weights than
for NNPDF weights.
This downwards trend persists even when more datasets of the same type are 
added to the prior.
Such a behaviour is not as marked in the NNPDF case, where instead the 
reduction of the effective number of replicas after the inclusion of the 
first dataset of a given type is only mild.\footnote{In
  Fig.~\ref{fig:neff_vs_data} the value of $N_{\rm eff}$ increases slightly 
  between datasets 6 and 7 for the GK case.
  When adding new measurements, the value of $N_{\rm eff}$ should
  always decrease (or remain constant) up to statistical fluctuations.
  These fluctuations are negligible when the number of starting replicas is 
  large enough, but not when one has only $N_{\rm eff}\simeq 20$ 
  replicas as in this specific case.
}

\begin{figure}[!t]
\center
\includegraphics[width=0.49\linewidth]{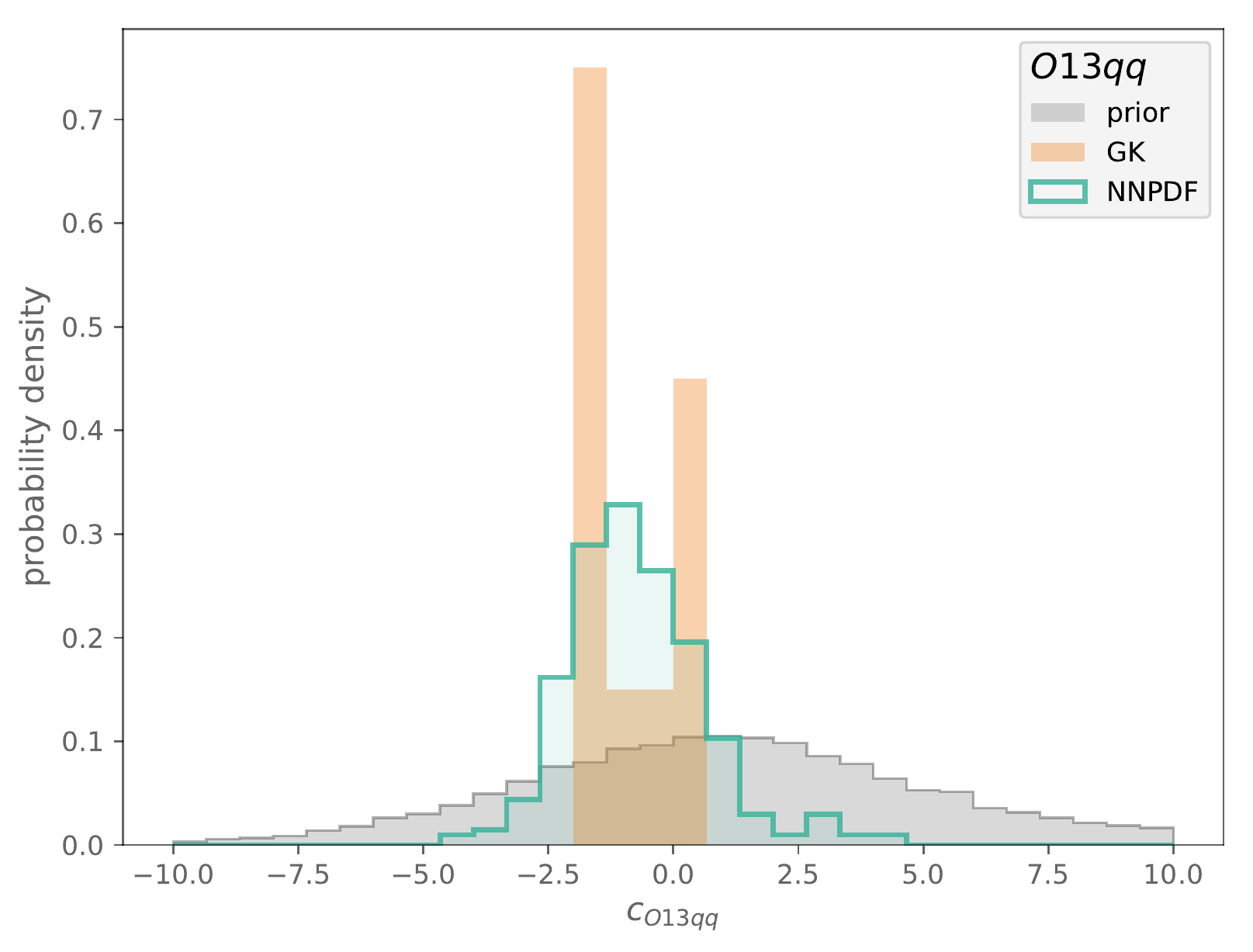}
\includegraphics[width=0.49\linewidth]{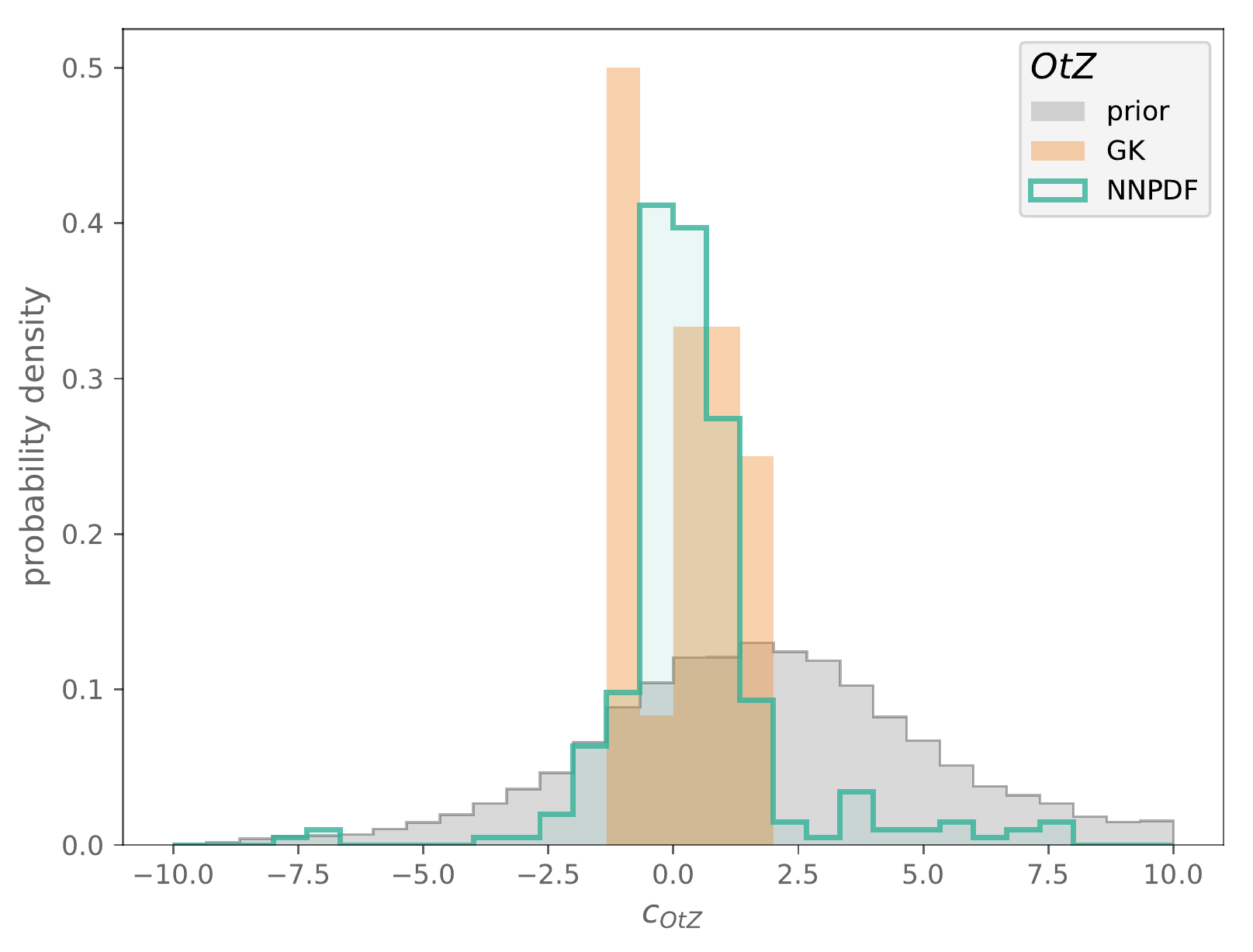}
\caption{\small Same as Fig.~\ref{fig:t_ch_coeff_dists} 
            for the distributions of the {\tt O13qq} and {\tt OtZ} operators 
            obtained from the prior and the reweighted sets with either the 
            NNPDF or the GK weights.
            The number of effective replicas in the
            two cases is $N_{\rm eff}\simeq 300$ and $N_{\rm eff}\simeq 20$, 
            respectively.
	    \label{fig:distr_O13qq_GK_full_t_channel}  }	
\end{figure}

Moreover, from Fig.~\ref{fig:neff_vs_data} one finds that, after 
the prior is reweighted with all the $t$-channel single-top cross sections 
in Table~\ref{eq:input_datasets2}, $N_{\rm eff}\simeq 300$ and 
$N_{\rm eff}\simeq 20$, respectively, in the NNDPF and GK cases.
In the latter case the effective number of replicas is simply too low for us to
trust the reweighted results.
The results of Fig.~\ref{fig:gk_vs_nnpdf_full_t_channel} should therefore
be interpreted with care.
Discrepancies between the NNPDF and GK cases could be explained as genuine 
differences between the corresponding weights, or else as large statistical
fluctuations due to finite-size effects.
This ambiguity is further illustrated in 
Fig.~\ref{fig:distr_O13qq_GK_full_t_channel}, 
where we compare the probability distribution of the {\tt O13qq} and {\tt OtZ}
operators in the prior and in the reweighted sets obtained both in the 
NNPDF and in the GK cases.
Differences between the reweighted NNPDF and GK distributions are apparent,
as is the fact that the GK distribution lacks sufficient statistics to
be reliable.

In order to understand whether the differences between the results obtained
with NNDPF and GK weights can be explained as finite-size effects, we need to 
compare results with a sufficiently large effective number of replicas.
A quick inspection of Fig.~\ref{fig:neff_vs_data} reveals that this can be 
achieved for both NNPDF and GK when only the first dataset is added.
In such a case, one would end up with $N_{\rm eff}\simeq 600$ and 
$N_{\rm eff}\simeq 250$ effective replicas, respectively.

\begin{figure}[!t]
\centering
\includegraphics[width=0.99\linewidth]{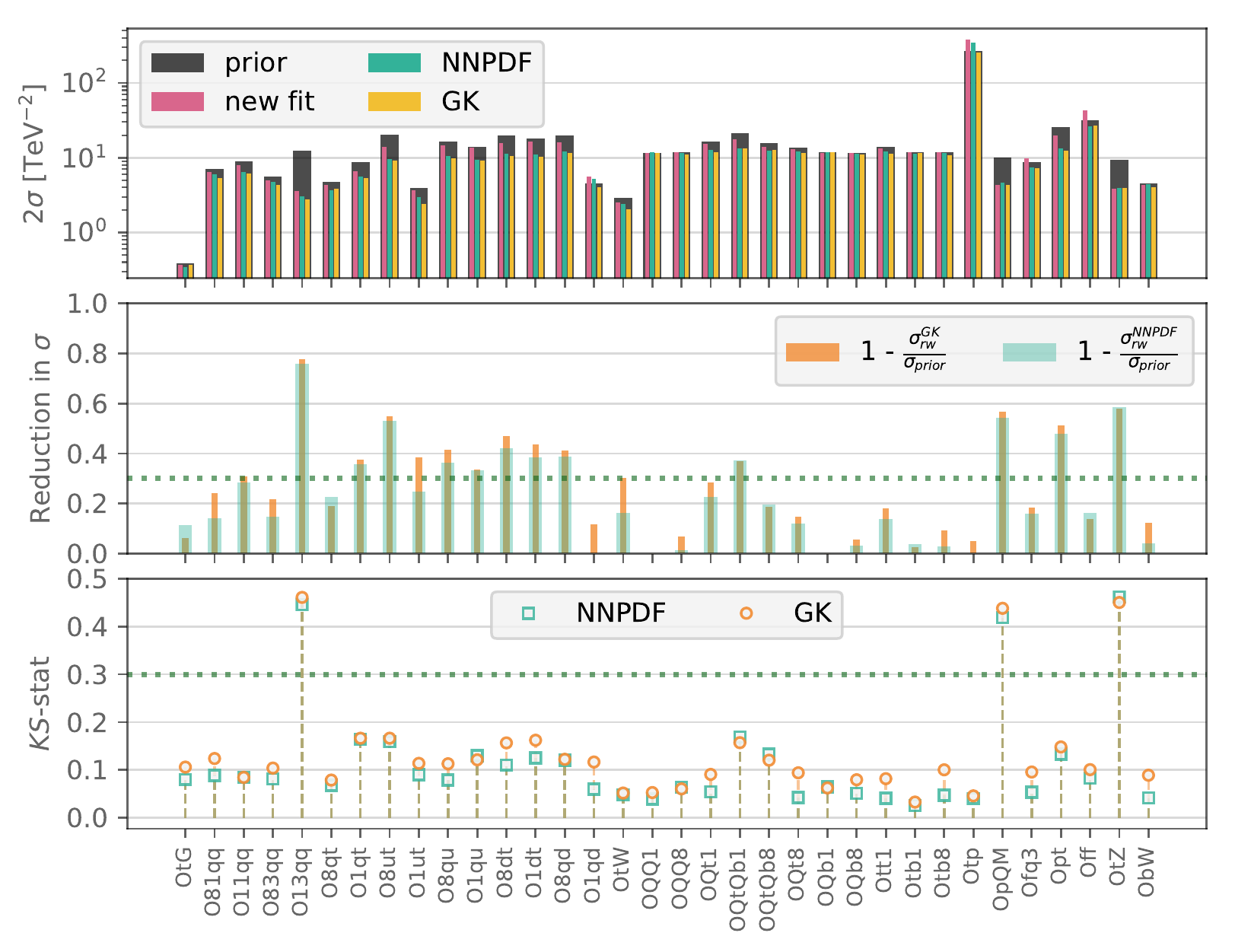}
\caption{\small Same as Fig.~\ref{fig:gk_vs_nnpdf_full_t_channel} now
      with only one $t$-channel single-top dataset
      (the differential $y^{t+\bar{t}}$ distributions
      from CMS at 8 TeV) added via reweighting 
\label{fig:gk_vs_nnpdf_1_t_channel}  }	
\end{figure}

In Fig.~\ref{fig:gk_vs_nnpdf_1_t_channel} we repeat the comparison shown in 
Fig.~\ref{fig:gk_vs_nnpdf_full_t_channel}, now obtained with the inclusion of 
the first single-top $t$-channel data point only.
Good agreement between the NNDPF and the GK reweighting results are 
now found for the three usual operators {\tt O13qq}, {\tt OpQM}, and {\tt OtZ}.
For other operators, residual differences are significantly reduced in 
comparison to Fig.~\ref{fig:gk_vs_nnpdf_full_t_channel}; most notably, the 
NNPDF and GK values of the KS statistic are now much more consistent between 
each other.

\begin{figure}[!t]
\centering
\includegraphics[width=0.49\linewidth]{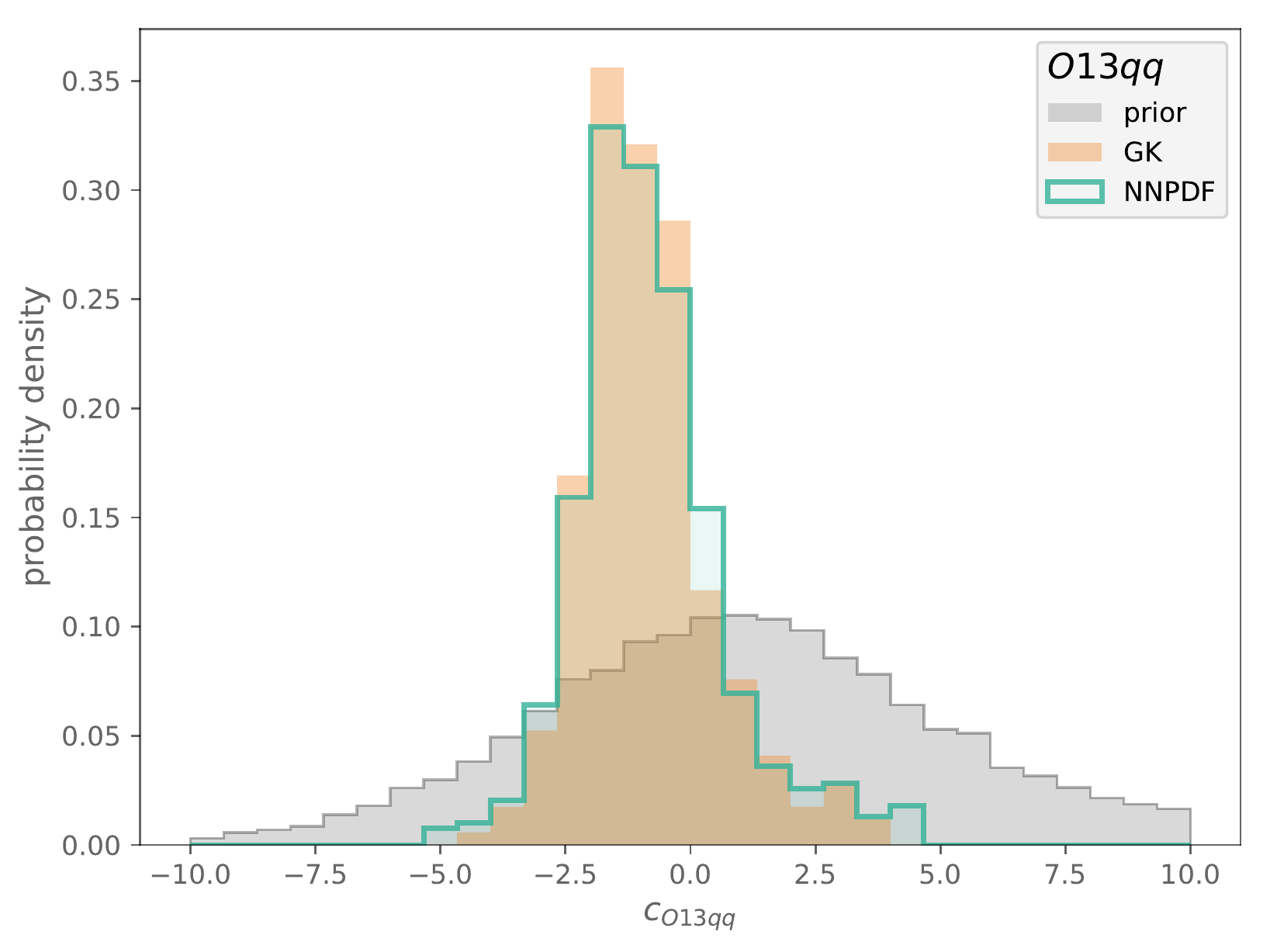}
\includegraphics[width=0.49\linewidth]{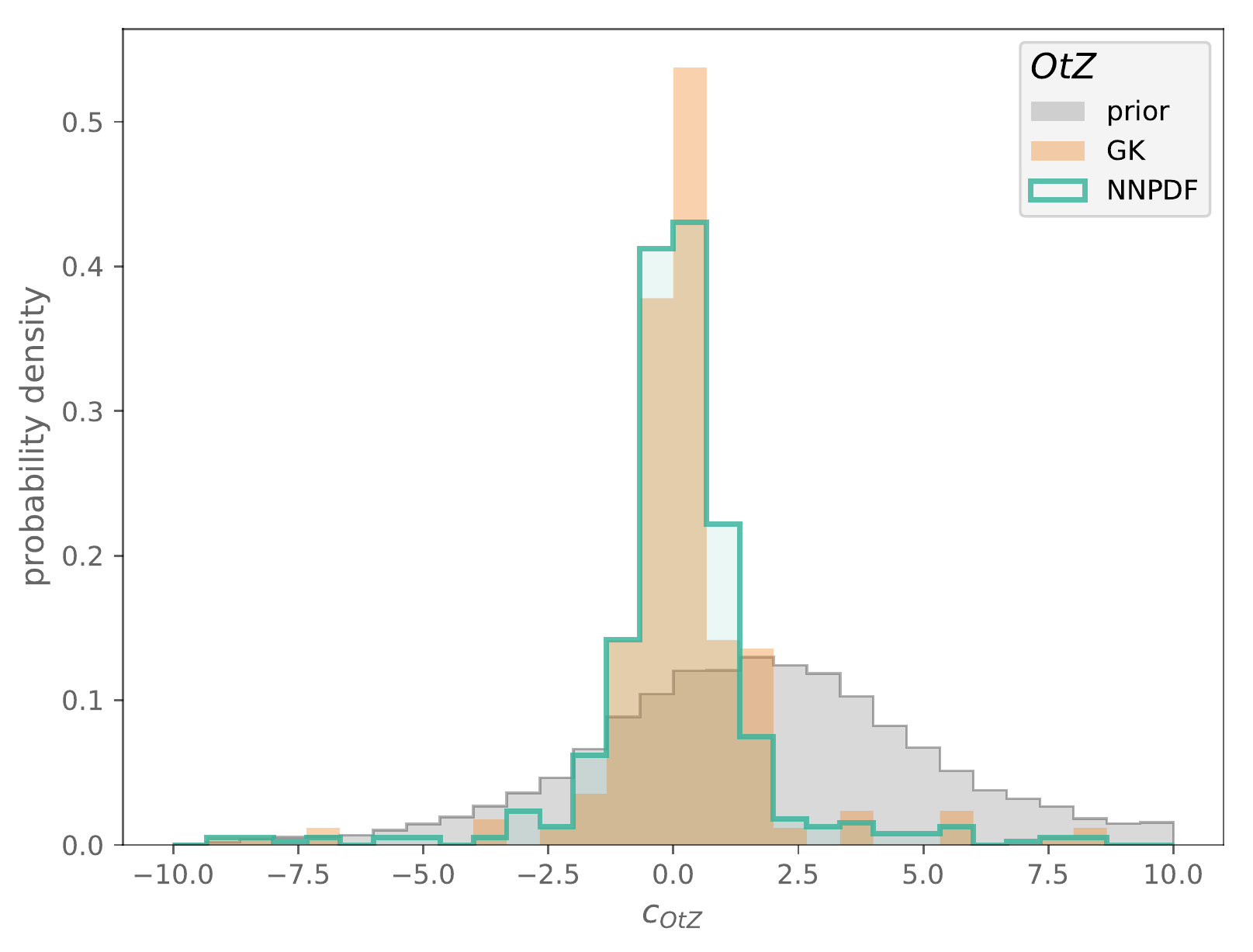}
\caption{\small Same as Fig.~\ref{fig:distr_O13qq_GK_full_t_channel}
      now for the case in which only one $t$-channel single-top dataset
      added via reweighting as opposed to the complete set of $t$-channel
      measurements, see also the results of Fig.~\ref{fig:gk_vs_nnpdf_1_t_channel}.
      The NNPDF results correspond to $N_{\rm eff}\simeq 600$
      while the GK ones to $N_{\rm eff}\simeq 250$.
      \label{fig:distr_O13qq_GK_1_t_channel}  }	
\end{figure}

Finally, in Fig.~\ref{fig:distr_O13qq_GK_1_t_channel} we repeat the comparison
between the prior, NNDPF and GK probability distributions of the Wilson
coefficients associated to the {\tt O13qq} and {\tt OtZ} operators when only
the first single-top $t$-channel data point is included.
Now that finite-size effects are under control, good agreement is found
between the NNPDF and GK reweighted distributions.
Such an agreement is consistent with Fig.~\ref{fig:gk_vs_nnpdf_1_t_channel}.

We find a very similar pattern of results if we repeat the above exercises 
with the single-top $s$-channel datasets.
We can therefore conclude that, based on the phenomenological exploration 
presented in this study, for those SMEFT operators that satisfy our selection 
requirements, and provided that the efficiency loss is not too severe (that is, 
the effective number of replicas is large enough), using either
the NNPDF or the GK weights leads to equivalent results.
Under the above conditions, these results agree with a corresponding new fit.
In general, however, GK weights appear to be rather less efficient than 
NNDPF weights.
This behaviour could easily lead to misleading results, unless one is careful 
in monitoring their dependence on the effective number of replicas.
In particular using the GK weights without ensuring that $N_{\rm eff}$ is 
sufficiently large might result in an overestimate of the impact that new 
measurements have on the SMEFT parameter space.


\subsection{Hybrid weights}

The rationale for comparing the results of the NNPDF and GK weights arises 
because in both cases there are studies that claim that one of the two is the 
correct expression based on formal arguments.
In principle it might be possible that there exists an intermediate expression
for the weights $\omega_k$ that interpolates between the NNPDF and GK weights 
while exhibiting a superior performance, although we are not aware
of any theoretical work supporting such a choice.
For completeness, here we investigate the performance of such hybrid
weights from the phenomenological point of view.

Specifically, we repeat the reweighting exercise that led to 
Fig.~\ref{fig:two_sigma_bounds_t_channel}, namely
including all the $t$-channel single
top production measurements.
Now, we use a one-parameter family of weights 
\be
\label{eq:weightsRWmod}
\omega_k^{(p)} 
\propto 
\lc \left(\chi^2_k \right)^{(n_{\rm dat}-1)/2}\rc^{1/p}
\,
\exp\left( -\chi^2_k/2\right) \, ,\quad k=1,\ldots, N_{\rm rep} \, ,
\ee
where $p$ is a parameter that interpolates between the NNPDF weights ($p=1$), 
Eq.~(2.4), and the GK weights ($p \to \infty$), Eq.~(4.1).
We have re-evaluated Figure 3.3 for different choices of $p$, and we have 
estimated the corresponding effective number of replicas.
This is what one finds for two intermediate values, $p=2$ and $p=3$:

\begin{table}[!h]
  \centering
\renewcommand{\arraystretch}{1.50}
\begin{tabular}{ccccc}
  \toprule
  $p$         & 1 (NNPDF) & 2   & 3  & $\infty$ (GK) \\ 
  \midrule
  $N_{\rm eff}$ & 306       & 115 & 68 & 22 \\
  \bottomrule
    \end{tabular}
\end{table}
%
\noindent Therefore there is no benefit in using the intermediate weights 
option: all values of $p$ decrease the efficiency of the reweighting procedure 
in comparison to the NNPDF case.

Furthermore, we have verified that, provided the effective number of replicas 
$N_{\rm eff}$ is large enough ($N_{\rm eff}\sim 100$), the results obtained 
with Eq.~(2) still reproduce the corresponding results of a new fit.
This is shown explicitly in Fig.~\ref{fig:two_sigma_bounds_t_channel_modn2}
below, the counterpart of Figure 3.3 but now using the hybrid weights in
Eq.~(\ref{eq:weightsRWmod}) with $p=2$.
As can be seen, for those operators for which the KS-statistic and
the reduction of uncertainties lie above the given threshold, the fit results 
are well reproduced by the reweighting with these hybrid weights.
However, for $p=2$, we end up with $N_{\rm eff}=115$ effective replicas, 
a value significantly smaller than the one obtained in the case of NNPDF
weights ($N_{\rm eff}=306$).

\begin{figure}[t]
\centering
\includegraphics[width=0.99\linewidth]{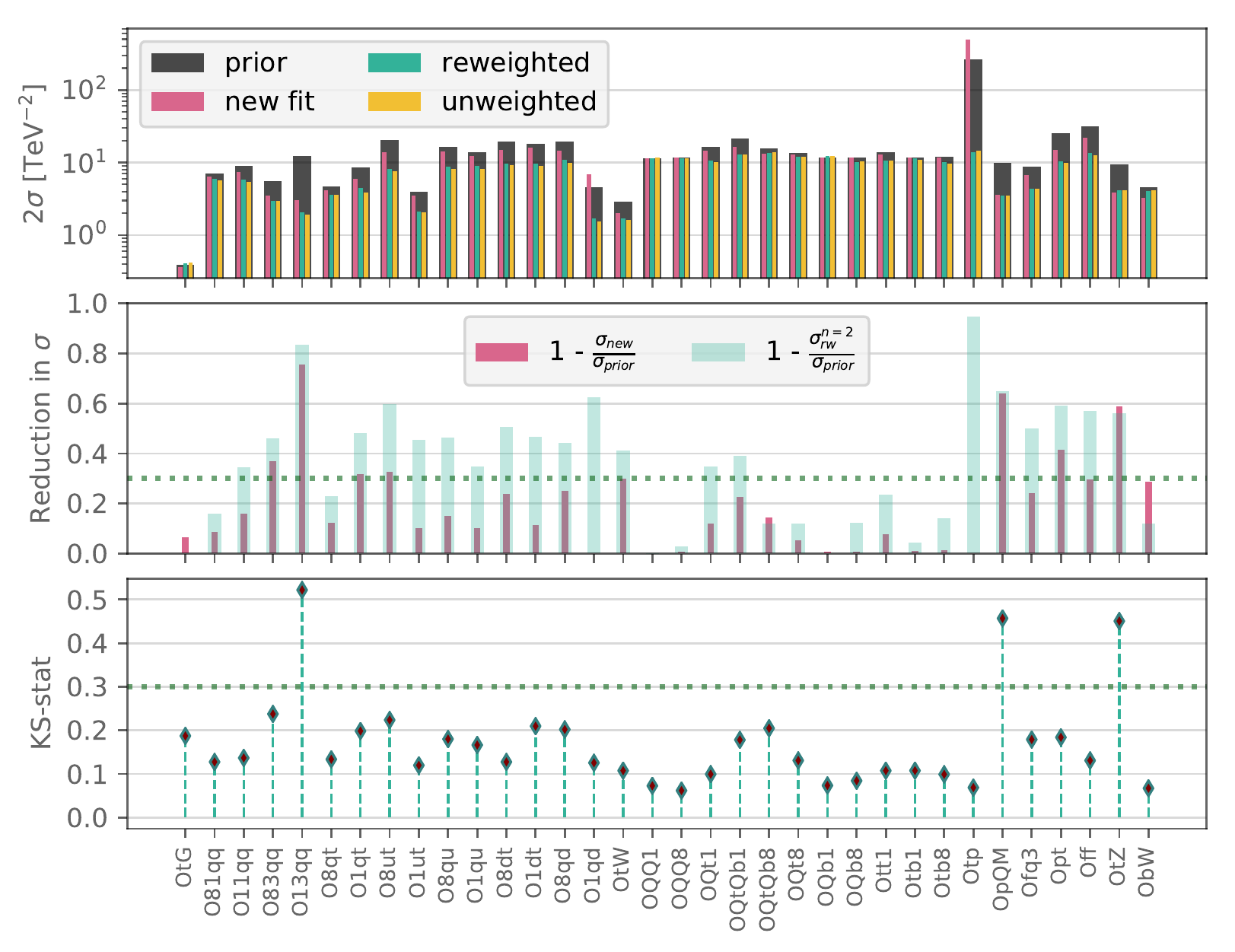}
\caption{\small Same as Fig.~\ref{fig:two_sigma_bounds_t_channel} but now using 
the hybrid weights defined in Eq.~(\ref{eq:weightsRWmod}) with $p=2$.
\label{fig:two_sigma_bounds_t_channel_modn2} }
\end{figure}

To summarise, we find that the hybrid weights of Eq.~(\ref{eq:weightsRWmod})
are equivalent to the NNPDF ones provided that the resulting 
$N_{\rm eff}$ is large enough, but also that they are less efficient since 
$N_{\rm eff}^{(p >1)} < N_{\rm eff}^{(p=1)}$.
Their use is therefore not advisable.


%% file: sec-summary.tex
\section{Summary}
\label{sec:sec-summary}

The lack of direct evidence for new physics at the LHC so far makes it crucial 
to develop indirect pathways to identify possible signatures of BSM dynamics
from precision measurements.
One of the most powerful frameworks to achieve this goal is provided by the 
SMEFT, which allows for a theoretically robust interpretation of LHC 
measurements.
Ensuring the model independence of this approach, however, requires us to 
efficiently explore the large parameter space of Wilson coefficients 
related to the SMEFT operators.
This can be both technically challenging and computationally expensive.

In this work we showed how Bayesian reweighting, an inference method 
widely used to assess the impact of new data in global determinations of PDFs, 
can be extended to constrain a Monte Carlo representation of the SMEFT 
parameter space upon the inclusion of new experimental input.
Reweighting consists of assigning a weight to each of the replicas that 
define the prior probability distribution.
These weights are computed as a function of the agreement (or lack thereof) 
between the prior and the new experimental measurements not included in it.
The method has two advantages in comparison to a new fit to an extended set 
of data: first, it is essentially instantaneous, and second, it can be carried 
out without needing access to the original SMEFT fitting code.
Using single-top production measurements from the LHC, we showed that, 
under well-defined conditions, the results obtained with reweighting
are equivalent to those obtained with a new fit to the extended set of data.

Nevertheless, the results obtained with the
reweighting method need to be considered with some care. 
First, it is necessary to verify that the efficiency loss of the reweighted 
sample is not so severe as to make the procedure unreliable. 
In practice, this requires that the effective number of replicas remains
sufficiently large.
Second, it is necessary to identify those operators for which
the results of reweighting are driven by a genuine physical effect,
and ignore those affected by large statistical 
fluctuations or other spurious effects.
In practice, this requires that the value of the KS statistic is sufficiently 
high. 
We therefore proposed a possible set of selection criteria to identify for 
which operators the outcome of the reweighting method is expected to coincide 
with that of a new fit.
We quantitatively based our criteria on the values of the KS statistic and of
the relative reduction factor for the uncertainties with respect to some 
threshold.

As a byproduct of this analysis, we also assessed the dependence of our results
upon the specific choice of weights, a topic which has been the cause of some 
discussions within the PDF community in recent years.
Specifically, we compared the performance of the NNPDF and GK weights,
proposed respectively in~\cite{Ball:2010gb,Ball:2011gg} and 
in~\cite{Giele:1998gw,Giele:2001mr}.
We found that, while the two methods lead to comparable results if the new 
data is not too constraining, the latter is far less efficient that the former,
in the sense that the effective number of replicas decreases much faster.
Our results provide further evidence in support of the use of the NNPDF rather 
than the GK expression for applications of Bayesian reweighting to Monte
Carlo fits, either in the PDF or in the SMEFT contexts.

The main limitation of the reweighting method is that it requires
a large number of starting replicas to be used reliably.
The problem is here somewhat more serious than in the PDF case, where an 
initial sample of $N_{\rm rep}=10^3$ replicas is usually sufficient for most
practical applications.
For instance, a prior of at least $\mathcal{O}\lp 10^5\rp$ replicas would be 
required for a simultaneous reweighting with all the $t$- and $s$-channel 
single-top data.
This happens because, in the SMEFT case, one is trying to simultaneously
constrain a large number of independent directions in the SMEFT parameter 
space, each corresponding to a different (combination of) operator(s).
However the efficiency loss should not represent a limitation to the 
applicability of the reweighting method, as in practice one usually wants to 
assess the impact of a single new measurement.
Of course, the reweighting method can only be applied
to explore directions in the parameter space 
that are already accessed in the prior set.
If new directions are expected to be opened by new data, for example
when measurements sensitive to different sectors of the SMEFT are included,
reweighting is not applicable and
a new fit would need to be carried out.

A {\tt Python} code that implements the reweighting formalism presented in 
this work and applies it to our SMEFiT analysis of the top quark 
sector is publicly available from 
\begin{center}
\url{https://github.com/juanrojochacon/SMEFiT}
\end{center}
together with the corresponding user documentation (briefly summarised in the Appendix).
In addition to the analysis code, we also make available three prior fits,
each of them made of $N_{\rm rep}=10^4$ Monte Carlo replicas.
The first fit does not include any $t$- and $s$-channel single-top quark 
production measurement, otherwise it is equivalent to the baseline fit of
Ref.~\cite{Hartland:2019bjb}.
It can be used to reproduce the results presented here.
The second fit is equivalent to the baseline fit presented 
in~\cite{Hartland:2019bjb}, but has $N_{\rm rep}=10^4$ Monte Carlo replicas
instead of $N_{\rm rep}=10^3$. 
It can be used to assess how the probability distribution 
of the top quark sector of the SMEFT is modified by the addition of new 
measurements via reweighting.
The final set is based only on inclusive top-quark pair production measurements.

%% file: sec-appendix.tex
\section{Code documentation}
\label{sec:codedocumentation}

The reweighting code in the publicly available {\tt GitHub} repository
consists of a single {\tt Python} script. 
It can be used straightforwardly by executing \texttt{SMEFiT\_rw\_unw.py} 
with {\tt Python3}. 
In order to run the code,
the following {\tt Python} packages need to be installed beforehand:

\begin{itemize}
	\setlength{\itemsep}{-1pt}
	\item {\tt numpy}
	\item {\tt tabulate}
	\item {\tt scipy} 
	\item {\tt matplotlib}
	\item {\tt os}
\end{itemize}

\subsection*{Code input}
\begin{figure}[!t]
\centering
\includegraphics[width=0.80\linewidth]{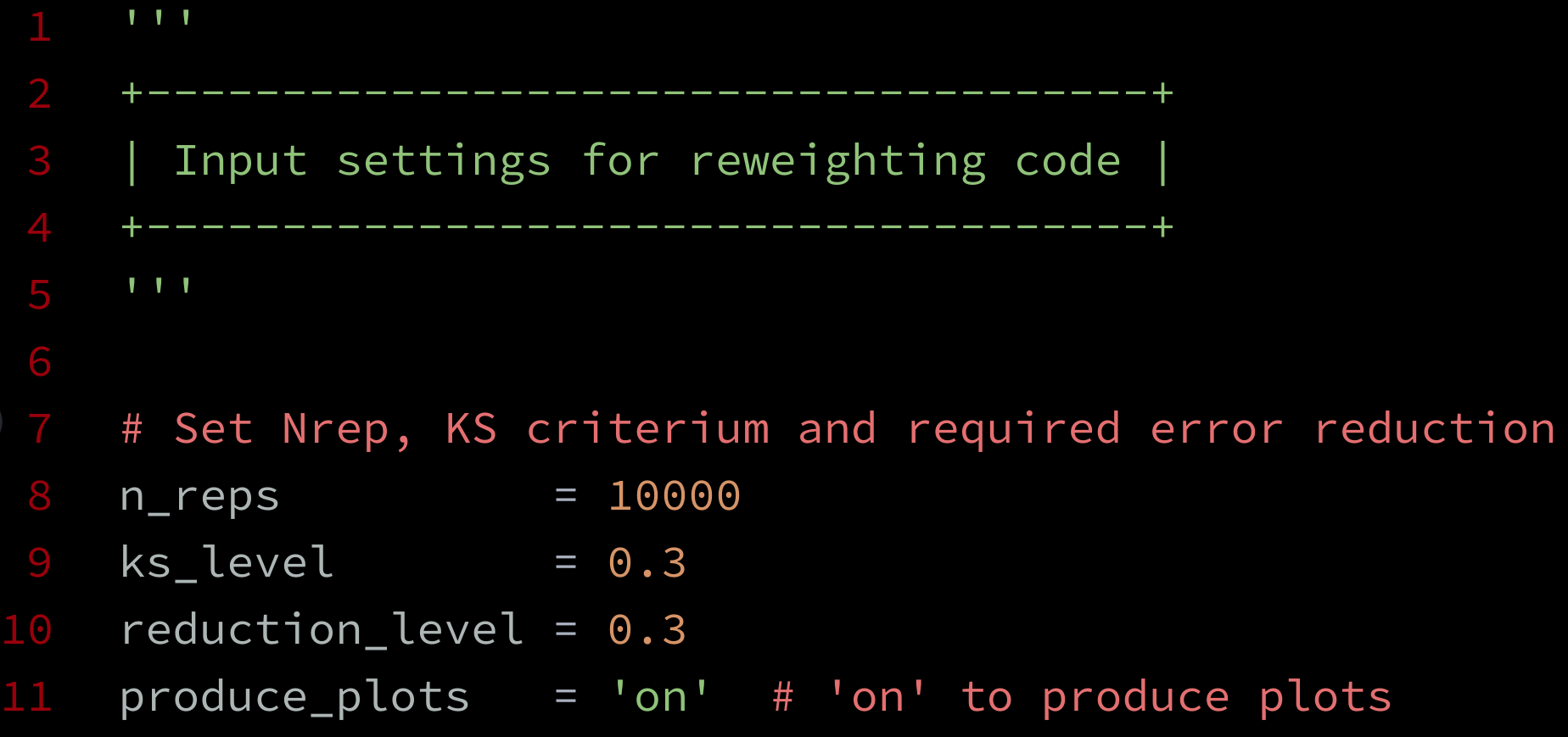}
\caption{\small A snapshot of the python script {\tt code\_input.py}.
         The input settings of the reweighting procedure can be defined here. 	
\label{fig:input_settings}}
\end{figure}

Besides the reweighting code file, there is also a second {\tt Python} 
file called \texttt{code\_input.py} that defines the input settings
to be used for the  reweighting procedure. 
In this file, the user is able to select the prior SMEFT Monte
Carlo analysis,
the datasets that he
wants to include in it by reweighting,
and the number of replicas that should
be used from it to this purpose.
By modifying this file, the user
can also specify the threshold values for the KS statistic and the error 
reduction factor that determine for which degrees of freedom
the reweighted results are reliable.
In Fig.~\ref{fig:input_settings}, a code snapshot of \texttt{code\_input.py} 
is shown. 

The following inputs will be required for the reweighting code to be executed:

\begin{itemize}
	
\item The Wilson coefficients that define the prior fit.

  Together with the code, we also release in the
  {\tt rw\_input\_data/Wilson\_coeffs/} folder the results
  of three different priors:
  {\tt all\_datasets},
  based on the full dataset of~\cite{Hartland:2019bjb},
  {\tt no\_single\_top}, with single top-quark
  production data excluded, and {\tt only\_ttbar}, based
  exclusively on top-quark pair-production measurements.

\item The replica-by-replica $\chi^2$ computed for the new data
  using the corresponding theory predictions based on a given prior
  SMEFT analysis.
  
  For illustration purposes, here we make available
  in the {\tt rw\_input\_data/chi2\_data/} folder
  the files {\tt t\_channel}, {\tt 1st\_t\_channel}, and
  {\tt s\_channel}, which are obtained from the prior
  set {\tt no\_single\_top} and correspond
  respectively to all $t$-channel single
  top-quark data, all $s$-channel single
  top-quark data, and only the first $t$-channel single
  top-quark measurement in Table~\ref{eq:input_datasets2}.

\item The Wilson coefficients determined from a new fit to the extended set 
  of data (if available, required only for validation).
  
  Here, also in the folder {\tt rw\_input\_data/Wilson\_coeffs/}, we
  provide the results of the  {\tt t\_channel}, {\tt 1st\_t\_channel}, and
  {\tt s\_channel} fits, which can then be directly
  compared to the corresponding reweighted results.

\end{itemize}

\subsection*{Code work-flow}

The code executes the reweighting and unweighting procedure through the 
following steps:
\begin{enumerate}
\setlength{\itemsep}{-1pt}
\item Load in the prior set.
\item Load in the replica-by-replica $\{\chi^2_k\}$ values 
      for the prior predictions.
\item Compute the weights $\omega_k$ for each replica.
\item Determine the Shannon entropy.
\item Obtain the reweighted set and construct the unweighted set.
\item Determine the KS statistic. 
\item Load in the new fit for validation (if available).
\item Determine the reduction of the uncertainty for the
  SMEFT degrees of freedom.
\item Obtain the operators that satisfy the
  selection criteria defined in 
      Sect.~\ref{sec:sec-validation}.
\item Save results to file and produce validation plots.
\end{enumerate}

\subsection*{Code output}
\begin{figure}[!t]
\includegraphics[width=0.99\linewidth]{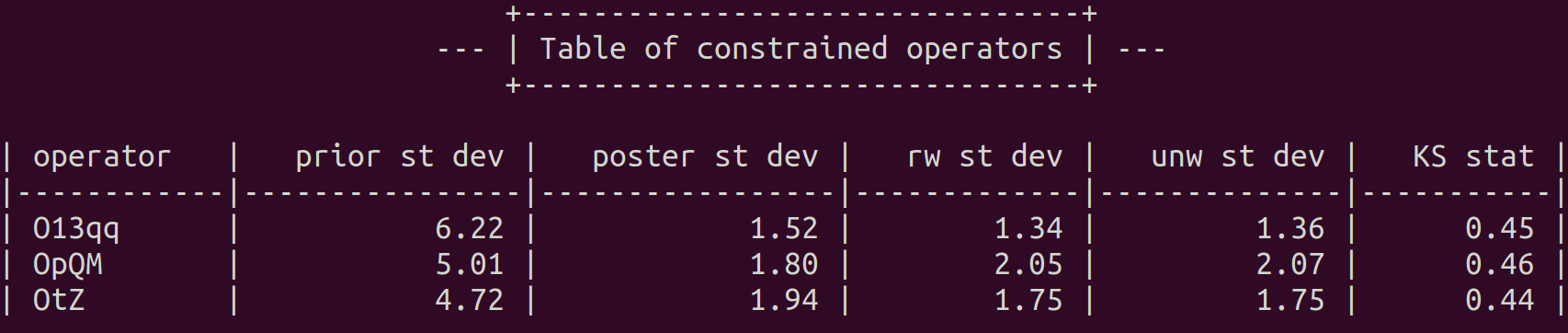}
\caption{\small A partial snapshot of the code output.
  The constrained SMEFT operators that satisfy the 
  reweighting
  selection criteria of Sect.~\ref{sec:sec-validation}
         are listed in a table with the value of their corresponding standard 
         deviation and KS statistic. 	
\label{fig:operator_table_terminal} }
\end{figure}

When the code is executed from a terminal, its output will be displayed
as in 
Fig.~\ref{fig:operator_table_terminal}. 
The following results will be saved in a new folder called {\tt rw\_output}:

\begin{itemize}

\item The plots of the $2\sigma$ bounds on the reweighted and unweighted Wilson 
  coefficients compared to the prior (and, if available, to the new fit
  for validation), 
  together with the associated uncertainty
  reduction factor and the value of the KS statistic.

\item The histograms for the
  distributions of the  Wilson coefficients
  for those operators that satisfy 
the selection criteria defined in Sect.~\ref{sec:sec-validation}.
 	
\item A text file {\tt unw\_coeffs.txt} with the
  results of the unweighted set of Wilson coefficients.
  In this file, the rows correspond to the replica number in the unweighted
  set, and the coefficients are separated in columns for each operator.
  
  Recall that the number of SMEFT operators in this unweighted set
  will be identical to that of the adopted prior.
\end{itemize}

Using this code,
the results presented in this paper can be easily reproduced by selecting
the same input settings as those adopted in the validation
exercises presented in Sect.~\ref{sec:sec-validation}.